\documentclass[letterpaper, 11pt]{article}
\usepackage[margin=1in]{geometry}
\usepackage{url}
\usepackage{color,soul}
\usepackage{tablefootnote}
\usepackage[official]{eurosym}
\usepackage[font=normalsize]{subfig}
\usepackage{lipsum}
\usepackage{booktabs,caption}
\usepackage[flushleft]{threeparttable}
\usepackage{graphicx}
\usepackage{dcolumn}
\usepackage{siunitx}
\usepackage{setspace}
\usepackage{titling} 
\usepackage{amsmath}
\usepackage{graphics}
\usepackage{adjustbox}
\usepackage{caption}
\usepackage{array}
\usepackage[capposition=top]{floatrow}
\usepackage{longtable}
\usepackage{multirow}

\usepackage{natbib}
 \bibpunct[, ]{(}{)}{,}{a}{}{,}%

\begin{document}
\title{\textbf{Context information increases revenue in ad auctions: \\ Evidence from a policy change}}
\author{
\textbf{S{\i}la Ada}\\
WU Vienna\\
\texttt{sila.ada@wu.ac.at}\\\\\
\textbf{Nadia Abou Nabout}\\
WU Vienna\\
\texttt{nadia.abounabout@wu.ac.at} \\\\\
\textbf{Elea McDonnell Feit}\\
Drexel University \\
\texttt{efeit@drexel.edu}\\
}

\pagenumbering{gobble}
\maketitle 
\vspace{3in}
\begin{center}
Working Paper\\
Please do not cite or distribute without permission. 
\end{center}

\newpage
\pagenumbering{arabic}

\begin{center}
\Large
\thetitle
\end{center}

\vspace{1in}

\begin{abstract}
Ad exchanges, i.e., platforms where real-time auctions for ad impressions take place, have developed sophisticated technology and data ecosystems to allow advertisers to target \emph{users}, yet advertisers may not know \emph{which sites} their ads appear on, i.e., the ad context. In practice, ad exchanges can require publishers to provide accurate ad placement information to ad buyers \textit{prior} to submitting their bids, allowing them to adjust their bids for ads at specific domains, subdomains or URLs. However, ad exchanges have historically been reluctant to disclose placement information due to fears that buyers will start buying ads only on the most desirable sites leaving inventory on other sites unsold and lowering average revenue. This paper explores the empirical effect of ad placement disclosure using a unique data set describing a change in context information provided by a major private European ad exchange. Analyzing this as a quasi-experiment using diff-in-diff, we find that average revenue per impression rose when more context information was provided. This shows that ad context information is important to ad buyers and that providing more context information will not lead to deconflation. The exception to this are sites which had a low number of buyers prior to the policy change; consistent with theory, these sites with thin markets do not show a rise in prices. Our analysis adds evidence that ad exchanges with reputable publishers, particularly smaller volume, high quality sites, should provide ad buyers with site placement information, which can be done at almost no cost.

\vspace{1cm}
\end{abstract} 

\newpage
\doublespacing
\section{Introduction}
\label{sec:intro}

\begin{figure}[b] 
\caption{Real-time bidding (RTB) markets match ad buyers with opportunities to advertise. }
\label{fig:RTB}
\centering
\includegraphics[width=\textwidth]{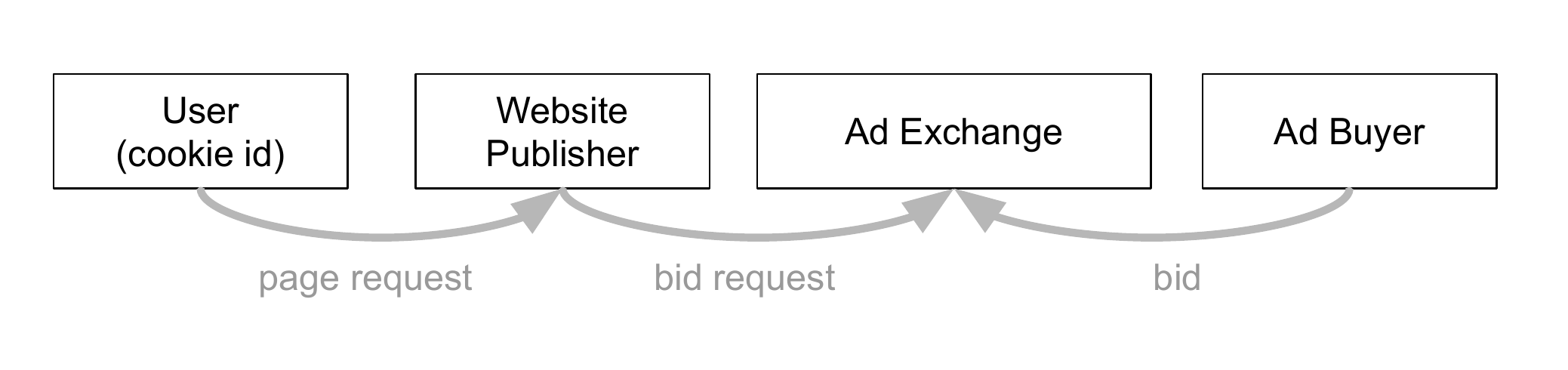}
\end{figure}

Digital display advertising has rapidly become popular among advertisers due to new targeting options as well as lower transaction costs enabled by ad tech. A central component of the industry is real-time bidding (RTB) markets for ad impressions (Figure \ref{fig:RTB}). When a user requests a page from a website, an impression (i.e., an opportunity to advertise to that user) becomes available. The site publisher can sell this impression on an RTB market by submitting a bid request to the ad exchange, which often includes a cookie ID identifying the user. The ad exchange subsequently broadcasts the bid request to potential ad buyers typically through intermediaries called demand side platforms (DSPs.) In response, ad buyers submit bids for the impression and the exchange sells the impression typically in a second-price auction. This entire process occurs within 400ms so that the ad loads almost instantaneously for the user. The publisher is paid the winning price less a commission to the exchange. An alternative to RTB is programmatic direct advertising, which allows ad buyers to pre-negotiate a guaranteed number of impressions at a particular site and are often processed using the same technology infrastructure as RTB. US programmatic digital display ad spending (including RTB and programmatic direct) is projected to reach \$68.47 billion representing 85\% of total digital display ad spending in 2020 \citep{emarketer19b}. Besides their economic importance, display advertising markets have been of interest to researchers \citep{goldfarb2011online, goldfarb2011privacy, reiley2012ad, lambrecht2013does, budak2014not, johnson2017consumer, berman2018, choi2018exchange, rafieian2020targeting, johnsonetal2020}, in part because they provide detailed tracking of ad transactions and user, ad buyer, and publisher behavior.

The display advertising industry has focused intensely on user targeting and there is a thriving market for third-party data describing the demographic characteristics and online behavior of each user cookie. In contrast, this paper provides evidence that display advertising buyers value information about the \emph{context} of the impression \textit{over and above} the information they already have about \textit{users}. Specifically, we show that when a private European ad exchange moved from providing bidders with no information about the website where an impression would appear to providing subdomain information (e.g., www.nytimes.com/section/business) the prices for ads increased for most sites, and overall for the ad exchange. Thus, buyers were bidding more for impressions when they knew more about the ad context indicating that they value the ad context and act on it. 

This finding has practical importance for the display advertising industry. While industry organizations like the Internet Advertising Bureau (IAB) have encouraged publishers to provide the exact URL where an impression will be placed in the bid request that the ad buyer receives prior to submitting their bid \citep{openRTB}, publishers have been slow to fully comply. For example, in January 2020, 15\% of bid requests passing through a large meta-exchange only included a high-level domain.\footnote{This data was provided by personal communication with a large meta exchange that aggregates about 400-600 billion bid requests each day across more than 150 supply-side platforms.} (Another 5\% included a subdomain and the remaining 80\% included the full URL.) When the high-level domain is included in the bid request, it often does not reflect the specific site where the ad will appear (e.g., when nbc.universal.com is listed as the domain in the bid request for any NBCU-owned website, even if the ad will appear at a different domain). To make matters worse, on the open exchanges some publishers provide outright fraudulent URLs in a practice known as ``domain spoofing" \citep{domainspoof}. Even though the authorized digital sellers (ads.txt) protocol from the IAB (Interactive Advertising Bureau) was touted to increase transparency and reduce fraud, publishers providing outright fraudulent URLs in the ads.txt file remains an issue \citep{Forbes20}. Thus, the site where an ad will appear is far from transparent to advertisers buying on the open exchanges today. Our study suggests that prices will rise when domain information is more transparent to ad buyers, benefiting ad buyers, reliable publishers, and exchanges. 

It was not obvious \emph{a priori} that context information would affect ad buyers' valuations for ads. When RTB was introduced to display advertising, the industry touted their capability to allow ad buyers to target specific users no matter what sites they were viewing (e.g., retargeting) and the industry has largely dismissed ad context as a poor proxy for the richer user targeting options available in digital advertising. However, there is some evidence in the academic literature that the website where an ad is placed can have an effect on the consumer's response to the ad \citep{goldfarb2011online, shamdasani2001location, lambrecht2013does, websitequal}. If an ad has a greater effect in a particular context, then we would expect ad buyers to value some site placements more than others. However, it was unknown, prior to the policy change, how much ad buyers would value context information relative to the rich user information already available. In addition, even if they valued it, it was unclear whether they would put that information to use given the complex decision environment for ad buyers in RTB.

While it was not clear whether ad buyers would respond to context information at all, it was also not clear how this information would affect the RTB market as a whole. Theoretically, providing ad buyers with the subdomain where the ad will appear allows them to put a more precise valuation on each impression that they bid on. If ad buyers all have the same (homogeneous) preferences over sites, then when ad buyers bid their valuations for individual sites rather than the average valuation of the bundle, prices will rise for some sites and fall for others. However, if ad buyers prefer different sites (heterogeneous preferences), then auction prices can rise uniformly across all sites. This somewhat surprising result has been shown theoretically \citep{hummel2016does} and empirically in an auction for used cars \citep{tadelis2015information}. But even under heterogeneous preferences, if too much information about an impression is provided, the market for a particular impression can become too thin \citep{levin2010online, hummel2016does} leading to a decline in prices. Using a quasi-experimental diff-in-diff analysis \citep{goldfarb2014conducting}, we find prices rose for most sites when more context information was provided, which is consistent with heterogeneous preferences among ad buyers. Our findings suggest that ad buyers value site placement information, that ad buyers have heterogeneous preferences for sites, and that providing context information moves the market to a higher point on the revenue curve for the platform and for most sites.  

To provide further evidence that prices rise because ad buyers have heterogeneous preferences for sites and bid higher on the sites they each prefer, we provide several mechanism checks. We show that when context information was available, a) slightly fewer ad buyers were winning impressions at each site (on average), b) prices rose for nearly all sites, and c) the distribution of winning bids became wider and has a thicker right tail. We also analyze a period when a single buyer was provided with ad placement information, and show that the buyer with the context information won more impressions, suggesting that the buyer was bidding higher on preferred sites. In total, these findings are consistent with theoretical predictions for second-price auctions where buyers have heterogeneous preferences for products and markets remain sufficiently thick.  

Finally, we investigate heterogeneous treatment effects and find that prices did not rise as much for sites that had fewer ad buyers prior to the policy change or were rated as providing low-quality advertising impressions (called non-premium sites) by industry experts. This is consistent with these sites being more at-risk of deconflation (i.e., developing thin markets,  \citet{levin2010online}) when their identity is revealed to buyers. We also find that sites providing a very small volume of premium impressions saw ad prices rise the most, suggesting that sites which provide ads to a ``niche" audience benefit the most from context disclosure. 

These findings are based on the observed outcomes in the exchange we study. This stands in contrast to much of the existing literature on ad auctions, which estimates structural models to bid-level data on auction outcomes and then reports counterfactual predictions for policy changes. For example, \citet{johnson2013impact} estimates a structural model from US-based ad auction data and uses this model to predict that both ad buyers and publishers are worse off when the platform introduces stricter privacy policies reducing user-targeting options. Similarly, \citet{lu2020} use a structural model to predict that by optimizing the level of information provided about users, an ad platform may improve its revenue. \citet{rafieian2020targeting} fit a structural model and use counterfactuals to predict the effect of limiting both user and context information on auction revenue and find that reducing context information affects revenue, but user information has a greater effect. Our approach is complementary to these studies in that it is based on careful analysis of the outcomes of an actual policy change, rather than structural assumptions that ad buyers and publishers behave rationally and consistent with auction theory.

We believe our findings will be even more important in the future, as privacy concerns lead to restrictions on the amount of user information that can be collected and shared with ad buyers. For example, Google recently announced efforts to limit the use of third-party cookies in their Chrome browser by making “disable third-party cookies” the standard setting \citep{chromecookies1, chromecoookies2}. Similarly, major ad-supported publishers like \emph{The New York Times} have moved to protect users by decreasing the amount of user information collected on their sites \citep{berjon2020} and some publishers have moved to eliminate user tracking all together \citep{edelman2020}. Thus, user targeting will be less straightforward in the future and industry experts predict that contextual targeting will become more relevant \citep{chromecoookies3, chromecoookies4}. This study provides direct, empirical evidence that context information is valuable to ad buyers.  

In the next section, we briefly review findings on information disclosure from the theoretical literature on auctions, which shows that prices may go up or down when more information is provided to all bidders, depending on market thickness and the heterogeneity in bidders' valuations. In the following section, we describe the institutional setting and policy change in more detail. We then analyze the empirical effect of information disclosure on ad prices using a diff-in-diff analysis, which shows that prices rose on average. Mechanism checks show that competition at each site decreased, but the right tail of the distribution of winning bids increased (i.e., winning bids started to spread out more) and nearly all sites saw an increase in revenue per impression. Following that, we compare the behavior of one buyer that received early access to the context information, compared to a synthetic control made of up of buyers without this information and find that this buyer won more auctions. This provides convergent evidence that ad buyers bid higher when they have context information for each impression, and markets thin-out a bit, but remain sufficiently thick for prices to rise. We then proceed with evidence for heterogeneous treatment effects related to 1) competition and 2) the size and quality of a site and conclude with a summary of our findings and a discussion of our study's implications.  

\section{Theoretical effect of information disclosure on auctions}
\label{sec:theory}

Theoretical predictions for how information disclosure affects ad auctions are mixed, with some researchers arguing that ad prices achieved in the auction should go up when more information is available about each impression \citep{hummel2016does} while others argue they should go down \citep{levin2010online}. These predictions depend critically on 1) the valuations of ad buyers for sites and 2) the number of bidders with positive valuations for each impression after the information is disclosed. If ad buyers all prefer the same sites, then prices will rise for the desirable sites and fall for the others. However, prices can rise for all sites if each ad buyer prefers different sites, i.e., buyers have heterogeneous preferences \citep{hummel2016does}, and markets are thick enough. 

For example, consider an ad exchange where there are two sites and bidder $i$'s valuation for an impression $j$ at site $k$ is composed of the value for the site $s_{ik}$ plus their value for remaining features of the impression $r_{ij}$ such as the cookie, time-of-day, etc. When context information is disclosed to all bidders, their bids are based on their valuation for each site. When context information is withheld, bidders are forced to base their bids on their average site valuation. For simplicity in this example, we assume that the bidders know that the two sites place equal numbers of impressions in the auction, thus the valuations are:
\begin{equation}
v_{ijk} = 
    \begin{cases}
      r_{ij} + s_{ik} & \text{if the site is disclosed}\\
      r_{ij} + (s_{i1} + s_{i2})/2 & \text{if the site is not disclosed}\\
    \end{cases}
\label{eq:valuations}
\end{equation}
For the examples that follow, we assume the valuations are independent and distributed as follows:
\begin{equation*}
r_{ij} \sim N(0, \omega) \hspace{0.25in}
s_{i1} \sim N(\mu - \delta/2, \sigma) \hspace{0.25in}
s_{i2} \sim N(\mu + \delta/2, \sigma)
\end{equation*}  
Here, $\omega$ represents variation in bidder's values for individual impressions, $\delta$ is the difference in average valuation between the two sites, and  $\sigma$ represents the variation across bidders in valuations for sites.

\begin{figure}
\caption{Effect of information disclosure on auction prices. Each dot represents the selling price of an impression in a simulated second-price auction where valuations are defined as in Equation \eqref{eq:valuations}. Average selling price is shown with horizontal lines. Panel (a) shows that when context information is disclosed to a large number of heterogeneous bidders, average price rises for both sites. Panel (b) shows that when there are only two bidders, disclosure causes prices to fall slightly. Panel (c) shows that when bidders homogeneously prefer site 2, prices fall for site 1 and rise for site 2. Panel (d) shows that when the residual value of the impression (due to cookie, time-of-day, etc.) dominates the site value, the context disclosure has little effect on the distribution of winning prices.}
\label{fig:sim_full}
\centering
\subfloat[]{
\includegraphics[width=0.48\textwidth, page=1]{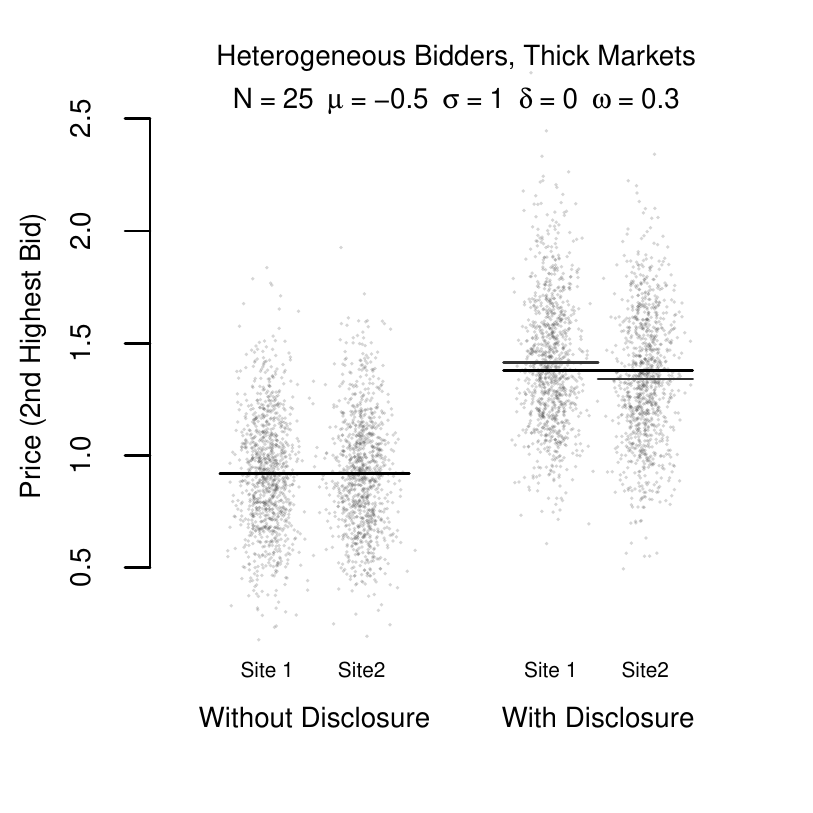} 
\label{fig:sim_het_thick}
}
\subfloat[]{
\includegraphics[width=0.48\textwidth, page=2]{Figures/simulation_graphs.pdf}
\label{fig:sim_het_thin}
}\\
\subfloat[]{
\includegraphics[width=0.48\textwidth, page=3]{Figures/simulation_graphs.pdf}
\label{fig:sim_homog}
}
\subfloat[]{
\includegraphics[width=0.48\textwidth, page=4]{Figures/simulation_graphs.pdf}
\label{fig:cookie_dominates}
} 
\end{figure}

When $\sigma$ is large and $\delta$ small, we have what \citet{tadelis2015information} refer to as ``heterogeneous bidders" or ``horizontal differentiation between sites".  When there are a sufficient number ($N$) of heterogeneous bidders, disclosure will increase auction prices for both sites because bidders bid more for the sites they uniquely prefer \citep{palfrey1983bundling, chakraborty1999bundling, hummel2016does, chen2018bundling}. Figure \ref{fig:sim_het_thick} illustrates this scenario showing simulated selling prices for 1,000 impressions where there are twenty-five bidders with heterogeneous preferences for sites ($\delta=0, \sigma=1$).  As can be seen from the horizontal lines in Figure \ref{fig:sim_het_thick}, average prices rise for both sites, although prices may rise more for one site than the other, depending on the realization of $s_{i1}$ and $s_{i2}$. Under context disclosure, the distribution of winning prices also has greater variance and a longer tail. As we show in the next section, the auction outcomes in the exchange we study are consistent with this scenario. 
The potential for prices to rise across the board when ad buyers have more information about impressions has also been shown analytically by \citet{hummel2016does}. They prove that average prices rise when there are at least four bidders with independently and symmetrically distributed valuations (so long as the distributions are not fat-tailed). Intuitively, this happens because the additional information leads to a better match between ad buyers and impressions. While this theoretical result has not been tested empirically for ad auctions, it has been observed in other types of auctions.  \citet{tadelis2015information} report on a field experiment where additional information on condition was provided to bidders in a used-car auction, resulting in an increase in prices for both high- and low-quality cars. They argue that the bidders have heterogeneous preferences for condition and that the information produces a better match between buyers and cars. 

Even in the case where there are heterogeneous preferences for sites ($\sigma$ large and $\delta$ small), if there are an insufficient number of bidders, prices may fall. This is illustrated in Figure \ref{fig:sim_het_thin}. which shows that average prices fall under information disclosure when there are $N=2$ bidders. Because there are so few bidders, some impressions have a second-highest bid of zero, shown by the large number of points with prices at zero for Site 2. (This does not happen for Site 1 because realizations of $s_{11}$ and $s_{21}$ are both high by chance.) \citet{levin2010online} describe phenomena like this in the context of user targeting and refer to it as an ``orphaned category". They argue that ad platforms should \emph{conflate} markets so that ``similar but distinct products are treated as identical in order to make markets thick or reduce cherry-picking." Not disclosing site information is one strategy for conflation, when markets are thin. 

Thus, Figures \ref{fig:sim_het_thick} and \ref{fig:sim_het_thin} show prices may rise or fall when relevant information is disclosed to bidders in an auction. For a given set of bidders and preferences, the literature on ad auctions concludes that the relationship between auction outcomes and information disclosed is concave with an intermediate amount of information (or equivalently bundling) producing the highest revenue \citep{rafieian2020targeting}. 

Context disclosure can lead to winners and losers among sites when ad buyers have homogeneous preferences for sites. We can simulate this scenario by setting $\delta$ higher and $\sigma$ lower. As Figure \ref{fig:sim_homog} shows, when bidders have homogeneous preferences for one site over the other, prices rise dramatically for the preferred site and fall for the other. This outcome may be revenue-neutral for the auction platform (depending on the mix of sites it represents), but it has a substantial effect on revenue for individual sites and may make publishers of less-preferred sites dissatisfied with the platform. The potential for this type of ``cherry picking" by ad buyers was a serious concern to the platform operator we study. However, the data from before and after the policy change suggest that ad buyers preferences for sites are largely heterogeneous (with some limitations, which we will discuss).

Finally, because it is possible that ad buyers do not value context information above-and-beyond cookie information, in Figure \ref{fig:cookie_dominates}, we simulate a case where bidders show more variation in their value for individual impressions and their cookies versus sites (i.e., $\sigma$ is small and $\omega$ is large). As the graph shows, when bidders place more importance on the cookie and other features of the impression than the context, then context disclosure has a modest effect on ad markets. 

Collectively, the simulations in Figure \ref{fig:sim_full} show that context disclosure may 1) uniformly raise prices for all sites, 2) uniformly lower prices, or 3) raise prices for some sites but not others depending on the number of bidders and their valuations for sites. It was also possible \emph{ex ante} that this policy change would have no effect at all because ad buyers may not behave rationally. The theoretical and structural literature on auctions relies on the assumption that bidders will maximize their expected value given available information. However, research on managerial decision making shows that managers are often risk-averse \citep{amihud1981risk} relying merely on historical performance patterns \citep{busenitz1997differences, little1970models}, such that they do not change their investment decisions when receiving better information \citep{lambert2007accounting}. Given the many potential targeting options available to online ad buyers, they may not have the time or incentive to adjust their bidding strategies for each site. Consequently, it is unclear whether ad buyers will put context information to use at all. 

To summarize, it is difficult to predict whether ad context information will affect auction outcomes for three reasons: 1) if site placement is not valued by ad buyers then the change will have no effect, 2) even if buyers value the context, they may not change their bidding strategy due to the complexity of the advertising environment, and 3) even when ad buyers are behaving optimally, the effect of information disclosure on auction outcomes is a complex function of buyer valuations and market thickness, and prices may fall or rise for particular sites or overall. Thus, it remains an empirical question how context information will affect RTB ad auction outcomes. Next, we describe the institutional setting where we study the effect of a change in context disclosure.

\section{Institutional setting}
\label{sec:data}

This paper investigates a change in the ad context disclosure policy at a major \emph{private} ad exchange in Europe. In a private exchange, a relatively small number of digital publishers agree to offer impressions to a pre-approved group of ad buyers through RTB. This makes private exchanges distinct from \emph{open} RTB exchanges like Google Ad Manager (formerly DoubleClick) where any ad buyer or publisher may participate and thousands do. While RTB began with the open exchanges, as concerns about transparency, fraud, and brand safety have grown, premium publishers including Hearst, Technorati, Conde Nast, CBS, NBCUniversal, IDG TechNetwork, The Weather Channel, and Vox have created private exchanges, which offer ads at a smaller set of reputable websites. Participating ad buyers and sites are vetted prior to bidding and the relatively small number of participants increases transparency and brand safety for both ad buyers and publishers. Software platforms for running ad exchanges such as Google AdMeld make it easy for small groups of publishers to build the necessary infrastructure to run an exchange. Sales on private ad exchanges are expected to exceed those on open exchanges in 2020 \citep{emarketer2020}.

\subsection{Policy change}
This private exchange offers us a unique opportunity to study the effect of a change in context information provided to ad buyers. Prior to April 2016, buyers (including advertisers themselves and intermediaries acting on behalf of multiple advertisers) on the exchange we study purchased ads without any knowledge of where the impression would appear. The only form of context-targeting available to ad buyers was buying ads on a  ``channel", where channels represented broad content categories like, ``news," ``automotive," or ``finance." Many ad buyers chose to place their ads on ``run of network," meaning their ads may appear on any of the participating publishers' websites, while still following any user targeting criteria the buyer has established.

In April 2016, a single large buyer was given access to context information in the bid request. Specifically, this buyer was given information about the subdomain where the ad would appear, e.g., nytimes.com/business. Throughout the analysis we use the term ``site" to refer to these subdomains, recognizing that some of these ``sites" are subdomains belonging to the same domain. In May 2016, after observing an aggregate rise in revenue when one buyer had site information, the exchange made the site for each impression available to all bidders. When the policy changed, ad buyers were notified by the ad exchange, often through personal phone calls. The specifics of how users determined bids specifically for sites varied by demand side platform (DSP); Figure \ref{fig:url} in the Appendix shows an example of how buyers were able to restrict their bids to particular subdomains in their bidding criteria. In addition to whitelisting or blacklisting sites, they could also use programmatic strategies to bid differently for different sites. 

The private exchange provides us with a well-controlled setting to study the effect of context information on ad buyers' valuations for ads. The participating publishers contractually agreed to sell \emph{all} their digital ad inventory exclusively through this exchange. Publishers could choose between RTB and programmatic direct sales and we observe all sales in both formats. Before the change, ad buyers knew that their ads would appear on one of the participating reputable sites, and not ``anywhere" as in open exchanges. Fraud by the publishers is not an issue in this setting. Finally, the policy change happened all-at-once, buyers could easily manage their buys to target specific sites, and all sales are fully-observed.

During this entire period, which was prior to the General Data Protection Policy (GDPR) coming into effect, ad buyers had access to the user cookie id for each impression and there was an active market for third-party data on past cookie behavior, so the policy change gives us insight into how ad buyers value context information over-and-above the rich user behavior data available at the time.   

The exchange hoped that this change toward greater transparency would make the exchange as a whole even more appealing to ad buyers and earn them certification as a brand-safe platform. However, the exchange had lengthy debates internally about the change. While some at the company were confident that revenue would rise for most or all sites, others were concerned that cherry-picking of the most desirable sites would lower revenues for less desirable sites, and potentially overall revenue for the platform (which gets a 2.5\% commission on all sales). As discussed in the previous section, these outcomes depend theoretically on the distribution of preferences for sites among the buyers, which was \emph{ex ante} unobservable to the exchange.

\subsection{Participating websites}

As discussed in the previous section, the theoretical effect of context disclosure in an ad auction depends critically on how the buyers value the websites where ads appear. The participating sites vary substantially in the types of content they provide and include one of the top 3 news sites in the country (according to SimilarWeb), one of the top 3 sports sites, and a variety of special interest and community sites similar to quora.com, webmd.com, allrecipes.com or zillow.com in the US.\footnote{Our agreement with the exchange precludes us from naming the sites or the country they operate in.} However, while some of the sites might be considered niche content, all of them are reputable and none would be considered extremist content or ``clickbait" (as you might find in the open exchanges).

The diff-in-diff analysis focuses on change in average revenue per impression for the 57 sites that participated in the market in both 2015 and 2016. To characterize the cross-sectional variation between sites, Table \ref{table:ss_supply} shows summary statistics on the supply of impressions and average revenue per impression for these sites during one week before the policy change. Across all sites, the mean revenue per thousand impressions (CPM) was \euro{0.88}. In addition, the average revenue per thousand impressions varies substantially across sites, with some sites selling for an average price as low as \euro{0.27} and others as high as \euro{2.16}. Prior to the policy change, ad buyers did not know which site the ad would be placed on when the impression was sold, so the variation in revenues across publishers in Table \ref{table:ss_supply} is due to differences in how ad buyers value the impressions based on the user-cookie and time of day. Table \ref{table:ss_supply} also shows that on average each site put 3,370,257 impressions into the auction during this week and sold 3,287,055, leaving an average of 2.5\% of impressions unsold for each site. The sites vary widely in the number of impressions that they provide to the market with a minimum of 737 impressions in this week versus a maximum over 104M impressions. The relatively high prices and low unsold inventory reflect the relatively high desirability of impressions on sites that participate in this private exchange. In this week prior to the policy change, the average daily buyers for each site, i.e., the average number of buyers who win impressions on a site each day is 57.66 (out of approximately 500 total buyers participating in the exchange).  
\begin{table}[ht]
\centering 
\caption{Summary statistics on impressions provided by individual websites during the second week of March 2016 (prior to the policy change).} 
\label{table:ss_supply}
\begin{threeparttable}
\begin{tabular}{@{\extracolsep{5pt}}lSSSSSS} 
\\[-1.8ex]\hline 
\hline \\[-1.8ex] 
Statistic & \multicolumn{1}{c}{Mean} & \multicolumn{1}{c}{St. Dev.} & \multicolumn{1}{c}{Min} & \multicolumn{1}{c}{Median} & \multicolumn{1}{c}{Max} \\ 
\hline \\[-1.8ex] 
Supply of impressions & {3,370,257} & {14,989,554} & {737} & {106,425} & {104,024,755} \\ 
Impressions sold &  {3,287,055} & {14,914,657} & {669} & {82,481} & {104,005,751} \\ 
Revenue per impression (CPM) &  0.88 & 0.39 & 0.27 & 0.81 & 2.16 \\ 
Average daily buyers &  57.66 & 37.79 & 7.57 & 50.57 & 159.14 \\
\hline \hline \\[-1.8ex] 
\end{tabular} 
\end{threeparttable}
\end{table}  
The primary data set we use for the analysis includes the number of impressions sold and the average selling price per impression for each website-buyer pair on each day. The data spans a period of seven months in 2016 that covers eleven weeks before the change (January - March 2016), four weeks where one ad buyer had access to placement information (April 2016), and twelve weeks where all ad buyers received placement information (May - July 2016). Similar data for the same months in the previous year is also available.\footnote{Note that we do not have access to bids or selling prices for individual impressions.}

\section{Context disclosure increased revenue per impression}
\label{sec:main_analysis}


Our goal is to identify how the policy change affected the average selling prices for impressions, i.e., the revenue per impression. As discussed in Section \ref{sec:theory}, if the buyers have heterogeneous preferences for the sites and there are a sufficient number of buyers for each site, then prices should rise overall. If there are sites that are undesirable to most buyers, such that the markets thin out, then prices may fall for some sites or overall. It is also possible that the market might not be affected at all, if buyers don't value the ad context, or if the transaction costs of customizing bids to specific sites are too high, or if buyers do not behave rationally. 

\subsection{Changes in revenue and supply of impressions} 

\begin{figure}[!ht]
\caption{Weekly average revenue per impression (CPM in \euro{}) and supply of impressions (millions) from January to July for 2015 and 2016. The first (second) red dashed line indicates the change to partial (full) disclosure.}%
\label{fig:time_trend}%
\begin{center}
\subfloat[Weekly average revenue per impressions]{\includegraphics[width=0.85\textwidth]{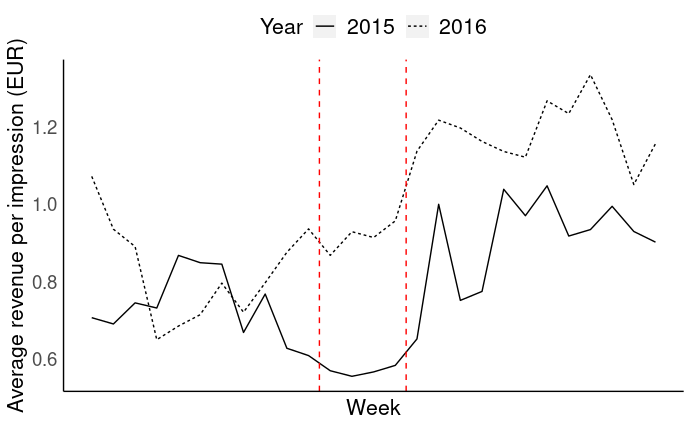}\label{fig:avg_price_of_imps}}%
\qquad
\subfloat[Weekly supply of impressions]{\includegraphics[width=0.85\textwidth]{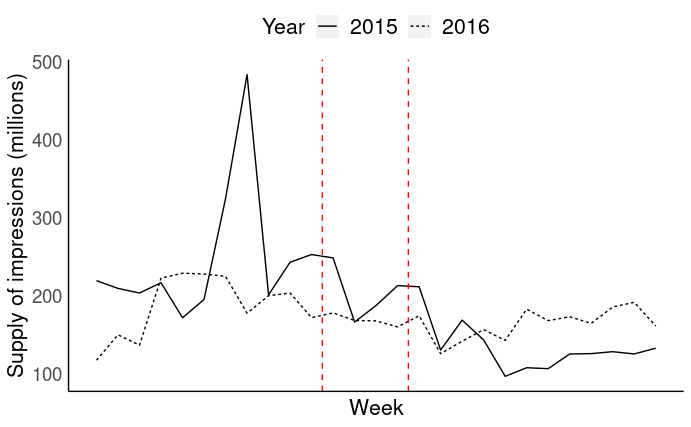}\label{fig:supply_of_imps}}%
\end{center}
\end{figure}

As an initial investigation of the data, Figure \ref{fig:avg_price_of_imps} plots the overall daily average revenue per impression (reported in \euro{} per thousand impressions, CPM) before and after the policy change. Figure \ref{fig:supply_of_imps} plots the total supply of impressions (sold and unsold) over the same period. The day when one buyer was given access to the subdomain for each impression in 2016 is indicated by the first vertical red dotted line and the day when all buyers were granted this context information is indicated by the second vertical dotted red line. Figure \ref{fig:avg_price_of_imps} shows that the average price per impression rose after the policy change. Total revenue on the exchange rose as well; it is substantially higher after disclosure in 2016 than it was in the same period the year prior (average of \euro{}155M per week versus \euro{}98M). It is also somewhat higher than it was earlier in 2016 prior to disclosure (average of \euro{}155M per week versus \euro{}134M). 

One notable feature in Figure \ref{fig:time_trend} is an unusual spike in supply from 2015-2-20 to 2015-2-27. We believe this is a data error as price seems to be largely unaffected, and report it for transparency.\footnote{As a robustness check, we re-do the diff-in-diff analysis excluding these observations and find that our substantive conclusions are unchanged.} 

More importantly, the full disclosure of context information coincides with the Spring and Summer where there is a lower supply of impressions and an increase in prices in both years. This observed decline in supply of impressions (and increase in prices) is consistent with seasonal patterns in web traffic, which tend to decline in the Summer. Thus, the policy change is confounded by the seasonal decrease in supply. In the next section, we do a formal diff-in-diff analysis with additional controls to account for this. 

\subsection{Diff-in-diff analysis} 
To show that the policy change increased revenue per impression for the platform, we regress the average revenue per impression for each site in each week on a 0-1 indicator for the policy change, a 0-1 indicator for the year and the interaction between those two, following the standard diff-in-diff approach. This regression is estimated using weekly data from January to July for both 2015 and 2016 and includes additional controls and site fixed effects. Thus our estimate of the effect of the policy change is the change in revenue per impression observed in 2016 above-and-beyond the change in revenue observed during the same period in 2015, i.e., the coefficient on the interaction between year 2016 and the policy change.

While the raw data is at the daily level, we summarized it at the weekly level to avoid having periods where a site did not sell any impressions. The dependant variable (average weekly revenue per impression for each website) is computed by dividing the total weekly revenue for each site by the total impressions that the site submitted to the RTB platform in that week including sold and unsold impressions. As our estimate of the treatment effect relies on a comparison of both years, we include only the 57 sites that sold impressions in both 2015 and 2016. This results in 3,058 site-week-year level observations, which is slightly fewer than 57 sites $\times$ 27 weeks $\times$ 2 years due to a few sites not selling impressions in the first few weeks of 2015. The regression is weighted by the number of impressions that each site placed in the RTB market in that week, so that the estimated effect of the policy change is an average change in price across all impressions in the RTB market. (This avoids overweighting sites that place fewer impressions in the auction, which, as we show later, saw larger increases in price.) 

To further account for fluctuations that might affect average revenue per impression, we include several control variables in the regression. First, we control for the total supply of impressions placed in the RTB market. Without controlling for the supply of impressions, we risk attributing the price change caused by the lower supply of impressions in the Spring and Summer (see Figure \ref{fig:supply_of_imps}) to the policy change instead. Second, we control for demand-side differences between sites by including the average daily buyers for each site prior to the policy change, which is a proxy for the total number of bidders competing for impressions at that site. Sites with more bidders should generally fetch higher prices. Third, as an additional control for overall demand for advertising in this market, we include the total national ad spending on digital advertising (sourced from a third-party data aggregator). Finally, we factor in site fixed effects that capture unobserved heterogeneity in sites.

Table \ref{tab:main_result}, Column 1 shows estimated coefficients for this regression. We report the increase in average revenue per impression for two periods: the period where one buyer had subdomain information in April, and the period when all buyers had subdomain information in May-July.\footnote{For those interested in how the treatment effect developed over time, we report effects separately for each month in the Appendix. Those results indicate that there was no ``learning" period for participants in the exchange: The observed treatment effect seems to have set in immediately.} The interaction terms \emph{Partial disclosure x Year16} and \emph{Full disclosure x Year16} are the key coefficients of interest, which show that prices increased by 10.8 EUR cents per thousand impressions during the period when one buyer had access to the subdomain information for each impression and by 15.4 EUR cents when all buyers had access to the subdomain information. This increase of 15.4 EUR cents for full disclosure is above-and-beyond the increase in average revenue per impression seen for the same months in 2015 (see the coefficient for \emph{Full Disclosure}) and the other controls. This substantial increase in prices represents the effect of moving this auction from ``channel" level context disclosure to subdomain level disclosure. 
\begin{table}[!htbp] 
\centering 
\caption{Diff-in-diff analysis of the change in average revenue per impression (\euro{} per thousand) due to context disclosure.} 
\label{tab:main_result} 
\begin{threeparttable}
\begin{tabular}{@{\extracolsep{5pt}}lD{.}{.}{-3} D{.}{.}{-3} D{.}{.}{-3} } 
\hline 
\hline \\[-1.5ex]
& \multicolumn{1}{c}{(1)} & \multicolumn{1}{c}{(2)} & \multicolumn{1}{c}{(3)}\\ 
& \multicolumn{1}{c}{Average revenue} & \multicolumn{1}{c}{Without controls} & \multicolumn{1}{c}{Placebo test} \\ 
& \multicolumn{1}{c}{per impression} & \multicolumn{1}{c}{} & \multicolumn{1}{c}{} \\ 
& \multicolumn{1}{c}{(CPM in \euro{})} & \multicolumn{1}{c}{} &\\
\hline \\[-1.5ex]
\textbf{Effect of policy change}  &  &  \\[1ex] 
Partial disclosure x Year16 & 0.108^{***} & 0.098^{***} & \\ 
  & (0.032) & (0.028) &  \\[1ex]  
Full disclosure x Year16 & 0.154^{***} & 0.111^{***} & \\ 
  & (0.039) & (0.043) &  \\[1ex]  
Placebo treatment x Year16 & & & -0.084\\
 & & & (0.062)\\[1ex]
\textbf{Baselines}  &  &  \\[1ex]  
Constant & 0.887^{***} & 0.929^{***} & 1.592^{***} \\ 
  & (0.147) & (0.017) & (0.236) \\[1ex]                  
Partial disclosure & -0.006 & 0.016 & \\ 
  & (0.020) & (0.022) &  \\ [1ex]  
Full disclosure & 0.182^{***} & 0.239^{***} &  \\ 
  & (0.035) & (0.038) & \\[1ex]  
Placebo treatment & & & 0.096\\
  & & & (0.060)  \\[1ex]
Year16 & 0.193^{**} & 0.234^{***} & 0.249^{***} \\
  & (0.027) & (0.027) & (0.046) \\[1ex]  
\textbf{Controls} &  &  \\[1ex]  
Supply of impressions  & -0.001^{**} & & -0.001^{***}\\ 
(millions) & (0.0004) & & (0.0004) \\[1ex]  
Average daily buyers  & 0.005 &  & -0.004^{***}  \\ 
(pre-treatment) & (0.001) & & (0.001) \\[1ex]  
Monthly ad spending & 0.0002 & & -0.0003 \\
& (0.0005) & & (0.001) \\[1ex]  
\hline \\[-1.5ex]
Site FE & \multicolumn{1}{c}{Yes} & \multicolumn{1}{c}{Yes} & \multicolumn{1}{c}{Yes} \\ 
N & \multicolumn{1}{c}{3058} & \multicolumn{1}{c}{3058} & \multicolumn{1}{c}{1234} \\ 
Adjusted R$^{2}$ & \multicolumn{1}{c}{0.7721} & \multicolumn{1}{c}{0.7639} & \multicolumn{1}{c}{0.7565} \\[1ex] 
\hline 
\hline 
\end{tabular} 
\begin{tablenotes}
\small
\item Notes: Standard errors, in parentheses, are clustered at the week level. $^{***}$Significant at the 1 percent level. $^{**}$Significant at the 5 percent level. $^{*}$Significant at the 10 percent level. A robustness check filtering out calendar week 6, 2015 when prices spiked showed substantively similar results.
\end{tablenotes}
\end{threeparttable}
\end{table} 

All control variables show the expected signs: average revenue per impression is lower when there are more impressions available in the market (i.e., supply is higher), sites with higher average daily buyers prior to the policy change (a proxy for the number of bidders, which is unobserved) have higher average revenue per impression, and average revenue per impression is higher when there is higher ad spending in the country where the exchange operates. Given that the supply of impressions and total digital ad spending are measured after the treatment, there is some potential for them to be endogeneous. However, the total spending on digital advertising in the country is determined across many markets including social media and search and is likely to be exogenous to this private exchange. The supply of impressions is potentially endogeneous if the publishers are moving impressions in and out of the market, but they are contractually obligated to sell all impressions in this exchange. The publishers can move impressions out of the auction and into programmatic direct deals, but we observe this and it did not happen (see Section \ref{sec:budgets} below). Thus the total supply of impressions is likely exogeneously determined by the number of users visiting the sites. However, a robustness check without the controls reported in Table \ref{tab:main_result} Column 2 shows effects that are somewhat attenuated without the controls, but still significant and substantial.

\begin{table}[!htbp]
\caption{Websites and ad buyers participating in the exchange before and after ad context disclosure.} 
\label{tab:continuity}
\centering
\begin{tabular}{l c c c c c}
\hline
\hline
& 2015 & 2016 Pre-change & 2016 Post-change & Total \\
& & (continuing + new) & (continuing + new) \\
\hline
Sites & 68 & 57 + 10 = 67 & 67 + 0 = 67 & 78\\ 
Buyers & 504 & 322 + 197 = 519 & 429 + 138 = 567 & 710\\
\hline
\hline
\end{tabular}
\end{table}
The diff-in-diff strategy for identifying the effect of disclosing context information to buyers relies mainly on the assumption that the seasonal pattern of revenue per impression is similar in 2015 and 2016 \citep{goldfarb2014conducting, datta2018changing}. While it is not possible to fully assess the parallel trends assumption from Figure \ref{fig:time_trend} as it does not include the other controls, the following evidence provides some confidence that revenue per impression in 2015 is a reasonable control. First, while display advertising markets can change quickly, this private market was quite stable. The sites in our data set are all well-established with stable traffic and inventory. The contracts between the exchange and the publishers can not be terminated easily or quickly. No sites dropped out of the private exchange during our observation window. Ad buyers, too, sign long-term contracts with the exchange, and buyer turnover in 2016 is fairly comparable to 2015 (see Table \ref{tab:continuity}). Second, we can see from Figure \ref{fig:supply_of_imps} that the observed supply decline in 2015 was slightly more pronounced than in 2016 suggesting that the change in average revenue per impression in year 2016 relative to year 2015 is a conservative estimate of the causal effect of the policy change. Finally, we carry out a so-called placebo treatment test using pre-period data from January to March 2016, estimating the effect of a placebo treatment starting at the mid-point of this pre-treatment data. The results are shown in Table \ref{tab:main_result}, Column 3, and reveal that we fail to reject the null-hypothesis of no treatment effect for our placebo treatment. Taken together, this evidence suggests that the parallel trend assumption is justified in this market.

Thus, the analysis shows that revenue per impression in the RTB auction increased when ad buyers were provided subdomain information for each impression. Since the policy change moved from disclosing nearly no information about where an ad would appear to disclosing the subdomain, this suggests that ad buyers value knowing the site where their ad will appear, even if they don't know precisely what content it will be placed next to.  Theory suggests that information disclosure will raise auction prices when 1) ad buyers have heterogeneous preferences for sites and 2) markets remain sufficiently thick for each site. In the next section, we provide additional evidence for this mechanism, but first we rule out an alternative explanation for the price increase. 

\subsection{Change in revenue per impression \textit{not} due to ad buyers shifting budgets into the RTB market} 
\label{sec:budgets}

\begin{table}[b]
\caption{Average monthly impressions (in millions) sold through RTB and programmatic direct in 2016 before and after the policy change} 
\label{tab:shift_prog}
\begin{tabular}{l D{.}{.}{1} D{.}{.}{-1} D{.}{.}{1} D{.}{.}{-1} D{.}{.}{1} D{.}{.}{-1}}
\hline
\hline
& \multicolumn{2}{c}{No disclosure} & \multicolumn{2}{c}{Partial disclosure} & \multicolumn{2}{c}{Full disclosure} \\
& \multicolumn{1}{c}{Imp.} & \% & \multicolumn{1}{c}{Imp.} & \% & \multicolumn{1}{c}{Imp.} & \% \\
\hline
Programmatic direct  & 1,320.1 & 59.4\% & 1,117.7 & 59.4\% & 1,144.8 & 59.4\%\\
RTB & 828.8 & 37.3\% & 744.3 & 39.6\% & 741.7 & 38.5\% \\
Unsold & 73.0 & 3.3\% & 19.2 & 1.0\% & 40.5 & 2.1\% \\
\hline
\hline
\end{tabular}
\end{table}
Prior to the policy change, ad buyers could not buy impressions at specific sites through the RTB market, but they could buy impressions at a specific site by making a programmatic direct deal with a specific site. These programmatic direct deals are often more expensive than prevailing prices in the RTB market. So, one explanation for why average revenue per impression increased is that ad buyers moved money out of programmatic direct deals and into the RTB market when it became possible to buy impressions at specific subdomains. However, we can rule this out, by looking at the proportion of impressions sold via programmatic direct and RTB in 2016. (The publishers were contractually obligated to sell all impressions through this exchange, so we observe how many impressions they sold through each mechanism.) Table \ref{tab:shift_prog} shows there is no obvious shift from programmatic direct to RTB associated with the policy change; the proportion of impressions sold via programmatic direct was nearly constant across the three periods. Thus, changes in buyers' bidding behavior in the RTB market is the more likely explanation for the rise in prices after the time of the policy change. Rather than using the RTB market as a replacement for purchases they previous made via programmatic direct, it appears that buyers were adjusting their bids in the RTB market. We provide more evidence of that in the next section.

\section{Mechanism checks}

Theoretically, information disclosure increases prices in an auction when there are a sufficient number of bidders who have heterogeneous preferences for the items. In this section, we provide additional evidence that this is the mechanism at play when context information was disclosed in this private exchange. Specifically, we show that 1) the market for each site was sufficiently thick after the policy change, 2) buyers appear to have been bidding higher for their preferred sites, and 3) most individual sites experienced an increase in revenue.   


\subsection{Market for impressions at each site remained sufficiently thick} 
\label{sec:stay_thick}

A necessary condition for prices to rise in the auction is that the number of bidders for each impression does not fall too low. To provide evidence that markets do not thin out, we investigate how the policy change affected the daily buyers for each site (i.e., the average number of winning bidders each day). Since we do not observe the individual bids, we can not say how many bidders were bidding on each impression, but the average number of daily buyers for each site gives us a proxy for the number of ad buyers who were submitting competitive bids for impressions at a given site. (The average daily buyers \emph{prior} to the policy change was used as a pre-treatment control variable in the diff-in-diff analysis; here we look at whether the average daily buyers \emph{changed} in response to the policy change.)  

\begin{figure}[!htbp]
\caption{Competitiveness of the auctions for each site as measured by the average daily buyers before and after the policy change (in 2016).}
\begin{center}
\includegraphics[width=0.8\textwidth]{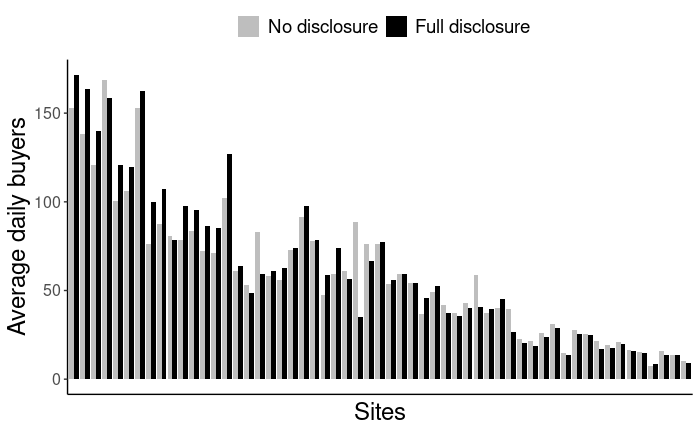}\\
\end{center}
\floatfoot{Note: Sites are sorted in order of weekly supply of impressions in the auction (highest to lowest).}
\label{fig:competitiveness}
\end{figure}
Figure \ref{fig:competitiveness} plots the average number of daily buyers for each site before and after the policy change and shows that there was little change in the number of buyers for each site. Thus the market for impressions at each site remained similarly competitive after the policy change. If at all, sites with thick markets seem to become slightly thicker, while sites with thin markets become slightly thinner. Yet, even the site with the fewest average daily buyers has an average of more than 8 buyers each day after the policy change. Therefore, the data suggests that most markets were sufficiently thick for prices to rise. 
\begin{table}[!htbp] 
\centering 
\caption{Diff-in-diff analysis of the change in average daily buyers.} 
\label{tab:change_ave_daily_bidders} 
\begin{threeparttable}
\begin{tabular}{@{\extracolsep{5pt}}lD{.}{.}{-3} D{.}{.}{-3} } 
\hline 
\hline \\[-1.5ex]
& \multicolumn{2}{c}{Average daily buyers} \\ 
\hline \\[-1.5ex]
\textbf{Effect of policy change}  &  &  \\
Partial disclosure x Year16 & -4.376 & (2.757) \\ 
Full disclosure x Year16 & -8.932 & (7.722)  \\[1ex]  
\textbf{Baselines}  &  &  \\
Constant & 43.565^{***} & (7.028) \\ 
Partial disclosure & 0.923 & (1.884)  \\  
Full disclosure & 28.686^{***} & (6.253)  \\
Year16 & 102.958^{***} & (2.302) \\[1ex]  
\textbf{Controls} &  &  \\
Supply of impressions (millions) & -0.028 & (0.023) \\ 
Monthly ad spending & 0.024 & (0.031) \\[1ex]  
\hline \\[-1.5ex]
Site FE & \multicolumn{2}{c}{Yes} \\ 
N & \multicolumn{2}{c}{3058} \\ 
Adjusted R$^{2}$ & \multicolumn{2}{c}{0.938} \\[1ex] 
\hline 
\hline 
\end{tabular} 
\begin{tablenotes}
\small
\item Notes: Standard errors, in parentheses, are clustered at the week level. $^{***}$Significant at the 1 percent level. $^{**}$Significant at the 5 percent level. $^{*}$Significant at the 10 percent level. 
\end{tablenotes}
\end{threeparttable}
\end{table} 

To provide additional evidence that markets remain thick after the policy change, we fit a regression for the total average daily buyers at each site as a function of the year and the policy change, using the same controls and site fixed effects as used in our diff-in-diff analysis. Table \ref{tab:change_ave_daily_bidders} shows a small, insignificant decrease in the average daily buyers when one buyer had access to the context information for each impression (\emph{Partial disclosure x Year16}). In the period when all ad buyers had access to this information, there is a slightly higher, but still insignificant decrease in the average daily buyers by -8.932 (\emph{Full disclosure x Year16}). Given the baseline of about 45 average daily buyers for each site, the number of buyers for each site decreased by about 18\% when everyone has context information, which is consistent with buyers bidding higher for the sites they each prefer leading to fewer average daily buyers for each site. However, markets remain sufficiently thick so that deconflation is not a concern.

\subsection{Distribution of winning bids shifts to the right} 
\label{sec:spread}

\begin{figure}[!ht]
\caption{Density plot of average price paid per impression before and after context disclosure.}%
\label{fig:density_plots}%
\begin{center}
\includegraphics[width=0.8\textwidth]{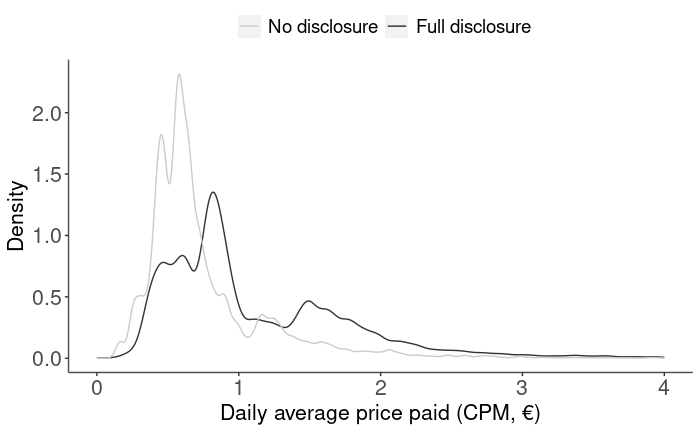}
\end{center}
\end{figure}

If buyers have heterogeneous preferences and are bidding more for their preferred sites when they have context information, then we should also see the right tail of the distribution of winning bids increase, consistent with the simulation results in Figure \ref{fig:sim_het_thick}. Indeed, this is the case. Figure \ref{fig:density_plots} plots the distribution of average prices paid by each buyer for each site and shows a distinct increase in the proportion of impressions selling for \euro{1.5-2.00} when ad buyers are provided with site placement information. This confirms that some buyers were bidding higher for some impressions. We do not see a similar shift in the distribution for these same time periods in 2015 (see Appendix Figure \ref{fig:dist_price_2015}).

\subsection{Most websites experienced an increase in revenue per impression} 
\label{sec:rise_for_all}

\begin{figure}[!ht]
\caption{Estimated change in revenue for individual sites, with 95\% confidence intervals. }%
\label{fig:price_change}%
\centering
\includegraphics[width=0.9\textwidth]{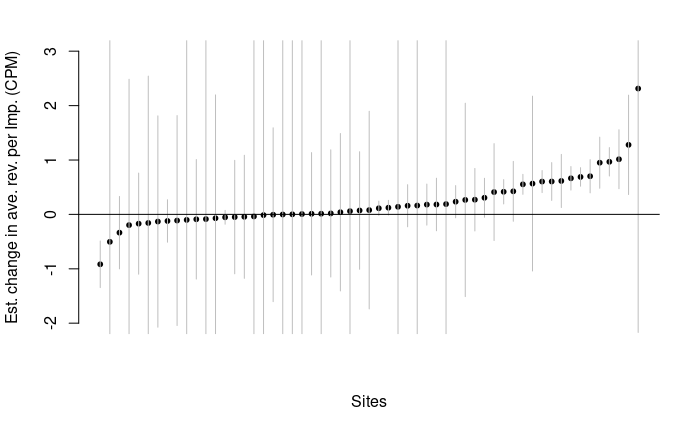}
\floatfoot{Note: The estimated change in revenue for an individual site is often estimated very imprecisely and in some cases the 95\% confidence intervals run off the plot.}
\end{figure}

If preferences are heterogeneous and markets remain thick, then most sites should see a rise in prices. Figure \ref{fig:price_change} plots the estimated effect of full context disclosure for individual sites and shows that revenue per impression rose for the majority of sites. These site-specific estimates are based on a regression with the same specification as our main diff-in-diff analysis in Table \ref{tab:main_result}, except that sites are interacted with the treatment indicators. There are few ``orphaned" sites; only one site shows a significant drop in revenue per impression.  The effect of context disclosure for most sites is either neutral or positive, with a few sites that gain substantially. This is consistent with some ad buyers having strong preferences for impressions at a particular site (above the information they already had about the user from the cookie.)  
  
Taken together, Sections \ref{sec:stay_thick}, \ref{sec:spread}, and \ref{sec:rise_for_all} paint a picture that is consistent with the changes expected for information disclosure in a competitive market where buyers have heterogeneous preferences. If each buyer is raising their bids for a different subset of sites, then average daily buyers for each site should fall slightly (as shown in Table \ref{tab:change_ave_daily_bidders}), the winning bids should have a longer right tail (Figure \ref{fig:density_plots}), and prices should rise for most sites (Figure \ref{fig:price_change}). 

\subsection{Partial information disclosure advantages the bidder with information}\label{partial}

As a final mechanism check, we investigate the effect of the policy change for the period where only the buyer who used a particular DSP was provided with ad placement information (April 2016). Theoretical research on information disclosure and bundling has focused on the cases like those illustrated in Figure \ref{fig:sim_full} where \emph{all} bidders have access to the same information and product offerings \citep{milgrom1982theory, eaton2005valuing, tadelis2015information, hummel2016does}. However, during a one-month period, the auction platform initially provided site information to one buyer only. To understand the expected effect of this partial disclosure, we make a brief departure to review another simulation showing the effect of disclosure to a single bidder.
 
\begin{figure}[!ht]
\centering
\includegraphics[width=0.7\textwidth, page=5]{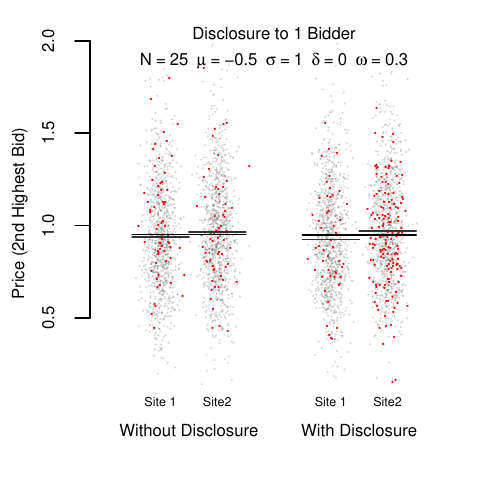}
\caption{Effect of information disclosure to a single bidder on auction prices. Each dot represents the outcome of a simulated auction and is colored red if the auction was won by the bidder who receives the additional information. Average resulting price is shown with horizontal lines.}
\label{fig:sim_partial}
\end{figure}

Specifically, we assume that one buyer has the site information and will bid their valuation under disclosure, while the other twenty-four buyers bid their valuation without disclosure (see equation \ref{eq:valuations}). The simulation otherwise follows the assumptions in Section \ref{sec:theory}. Figure \ref{fig:sim_partial} shows simulated winning prices for no disclosure versus partial disclosure with the auctions won by the first bidder colored red. In this scenario, when information is disclosed to just one bidder, that buyer's bids are higher for their preferred site, resulting in the treated bidder winning more often. In this example, the treated bidder wins 6.3\% of impressions without disclosure and 7.2\% when they have context information. The amount by which the treated bidder wins more depends on the bidders individual preferences for sites; in this simulation, the treated bidder had a fairly high valuation for Site 2.

However, the average winning prices do not necessarily change substantially under partial disclosure; whether that bidder pays more or less depends on the mix of sites that the treated buyer purchases, which depends on the valuations of all the bidders. Thus, when there is a large number of bidders who are heterogeneous in their valuations, partial disclosure theoretically results in the \emph{treated bidder winning more}.

To understand what happened in practice at the private exchange, we analyze the behavior of the bidder who obtained exclusive access to site placement information in April 2016 relative to other bidders. This buyer was a DSP bidding on behalf of several advertisers. Prior to the policy change, the treated bidder paid higher prices than the average of all other bidders (compare the first column in Table \ref{tab:predictor_means} to Column 4). 

To construct a counterfactual for what would have happened if this buyer had not gotten placement information, we use a synthetic control analysis \citep{abadie2003economic, abadie2010synthetic} to construct a counterfactual buyer that resembles the treated buyer during the pre-intervention period. The counterfactual synthetic control is a convex combination of untreated buyers that matches as closely as possible on several pre-treatment covariates known as ``predictors" in the synthetic control literature. The weights that define the control buyer are chosen such that the counterfactual buyer's predictors approximate the treated buyer's during the period prior to the policy change, week-by-week. Then, the constructed synthetic buyer is used to estimate a causal counterfactual for how the treated buyer would have behaved if not provided with placement information. We analyze both the winning price and impressions won as dependent variables and construct separate synthetic controls for each. (Note that average winning price is the same as average revenue per impression but at the buyer-week level.) Technically, the impressions won and prices paid by buyers who did not have access to this information may have been affected somewhat since they are participating in the auction with the treated bidder. However, since our treated buyer represents less than 2\% of impressions sold, the control buyers would have only been affected by a small amount.

The synthetic control is matched on the following predictors: (1) number of impressions won in each genre in each pre-treatment week, (2) average of price paid in each genre in each week, (3) total number of impressions won in each week and (4) average price paid in each week. Genres were utilized to create the predictors, since ad buyers were able to target channels or users based on behavioral information prior to the policy change. The covariates are created based on February-July in 2015, as well as February-March 2016. In constructing the synthetic control, daily observations with substantially higher prices (e.g., \euro{10} CPM) or lower volumes (\textless500 impressions in a day) were filtered out. These unusual observations are likely due to highly targeted buys that are not representative of the types of prices paid by the treated buyer. They represent only 0.8\% of impressions. The core identifying assumption is that these pre-treatment covariates represent the key ways in which the treated buyer is different than the untreated. 

Table \ref{tab:predictor_means} reports the summary of the covariates used in the construction of the synthetic buyer and compares them to the treated buyer, which are by construction largely similar.\footnote{The weight associated with the predictor ``number of impressions won on Community \& Forums sites" is very small, which indicates that it does not have predicting power for either dependent variables.} Furthermore, buyers in the control group that are picked by the algorithm are mainly the same for both independent variables and the highest weights are assigned to other intermediaries bidding on behalf of multiple advertisers (similar to the treated buyer).
\begin{table}[htbp!]
\centering 
\caption{Descriptive statistics for pre-treatment behaviors used to construct the synthetic control (average of predictors over pre-treatment weeks).} 
\label{tab:predictor_means} 
\begin{threeparttable}
\begin{tabular}{@{\extracolsep{5pt}} lcccc} 
\\[-1.8ex]\hline 
\hline \\[-1.8ex] 
& & {Synthetic} & {Synthetic} & {Mean}\\ 
& {Treated} & Control & Control & {Other Buyers}\\
& & for Price & for Impressions \\
\hline \\[-1.8ex] 
Impressions purchased\\
\hspace{0.25in} Overall& 1,126,804 & 1,412,513 & 1,366,452 & 1,176,408 \\ 
\hspace{0.25in} on Community \& Forums & 70,926 & 49,306 & 41,962 & 43,349 \\ 
\hspace{0.25in} on  General interest & 992,351 & 1,266,406 & 1,231,817 & 1,065,867 \\ 
\hspace{0.25in} on  Health & 4,147 & 5,328 & 4,409 & 4,834 \\ 
\hspace{0.25in} on  Special interest & 3,603 & 1,714 & 1,687 & 6,771 \\ 
\hspace{0.25in} on  Sports & 55,778 & 89,759 & 86,577 & 55,587 \\ 
Price  paid \\
\hspace{0.25in} Overall & 3.46 & 3.22 & 3.05 & 0.40 \\ 
\hspace{0.25in} on  Community \& Forums & 3.86 & 3.62 & 3.09 & 0.33 \\ 
\hspace{0.25in} on  General interest & 3.43 & 2.59 & 2.83 & 0.35 \\ 
\hspace{0.25in} on  Health  & 2.30 & 2.39 & 1.44 & 0.14 \\ 
\hspace{0.25in} on  Special interest & 1.14 & 0.94 & 0.75 & 0.13 \\ 
\hspace{0.25in} on  Sports & 3.69 & 3.42 & 2.96 & 0.32 \\ 
\hline
\hline\\[-1.8ex] 
\end{tabular} 
\end{threeparttable}
\end{table} 

\begin{figure}[!htbp]
\caption{Comparison of average winning price and number of impressions won for treated buyer versus synthetic control.} \label{fig:trends}%
\subfloat[Average winning prices (CPM in \euro{})]{
\includegraphics[width=0.85\textwidth]{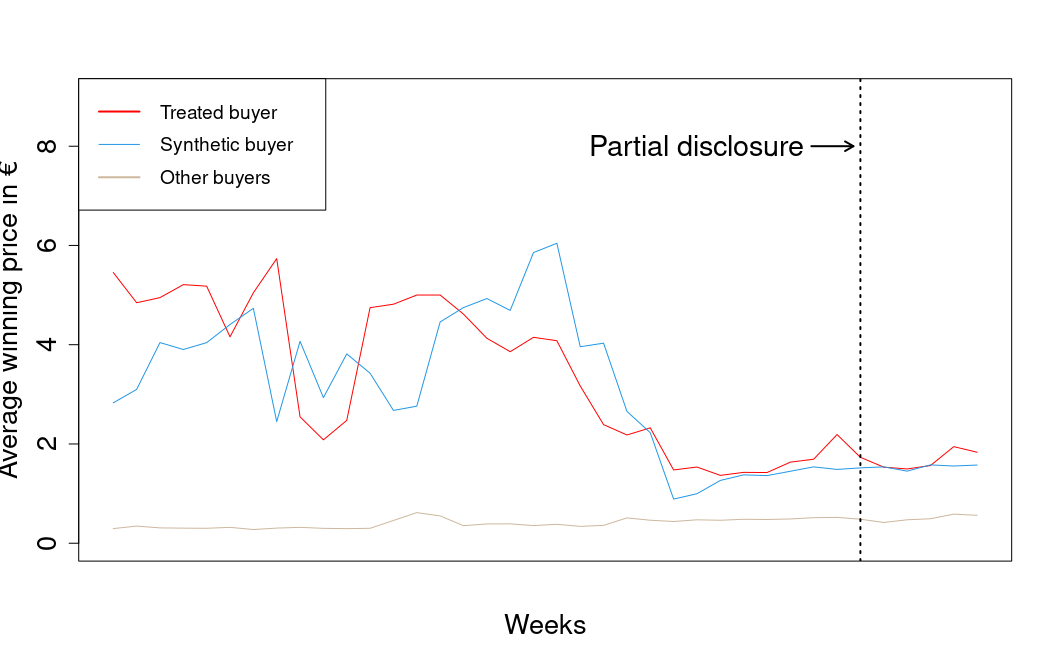}
\label{trends_price}
}\\
\subfloat[Number of impressions won]{%
\includegraphics[width=0.85\textwidth]{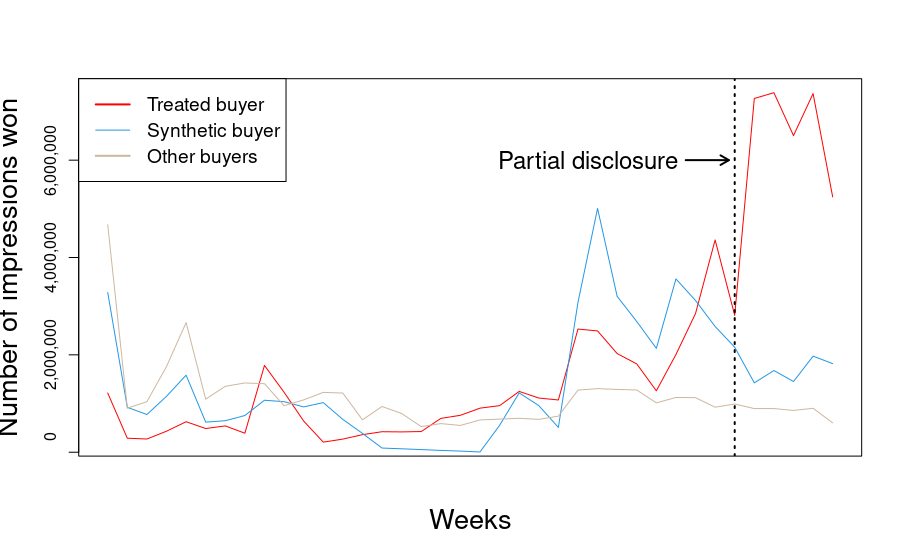}
\label{trends_imps}
}
\floatfoot{Note: Pre-treatment time series includes February - July 2015 and February - March 2016 and is thus discontinuous in time.}
\end{figure}

Figure \ref{trends_price} shows that the trajectory of the synthetic buyer's average winning price closely follows the treated buyer's price, which suggests that the synthetic buyer nicely mimics the treated buyer prior to policy change. Consistent with the simulation in Figure \ref{fig:sim_partial}, the additional placement information does not affect the average winning price for the treated bidder, as can be seen by comparing the treated bidder to the synthetic control in Figure \ref{trends_price} after the policy change.

However, the simulation in Figure \ref{fig:sim_partial} suggests that the number of impressions won by the treated bidder should be higher when provided with placement information, so we compare the number of impressions won for the treated and synthetic control bidders in Figure \ref{trends_imps}. To the left of the vertical line, the number of impressions won is similar for the treated and synthetic control bidder suggesting that the algorithm was able to find a comparable synthetic buyer. After the policy change, we see a large gap between the treated and synthetic buyer in the number of impressions won. While impressions won by the treated ad buyer increased substantially after the policy change, they remained flat for the synthetic control. The statistical significance of these results are confirmed by placebo tests reported in Figure \ref{fig:gaps} in the Appendix.

In this section, we have provided convergent evidence that ad buyers have heterogeneous preferences for sites and that the markets were sufficiently competitive, which resulted in the overall rise in prices with context disclosure. In the next section, we explore heterogeneous treatment effects. 

\section{Heterogeneous treatment effects}

Up to this point, the analysis has focused on how providing context transparency affected the prices for impressions on average. The analysis suggests that buyers value context information and buyers have heterogeneous preferences for sites, which leads to revenue gains for all sites. Next we turn to the practical question of whether certain types of sites benefited more from this policy change. First, we show that sites with more buyers prior to the policy change, i.e., those with thicker markets, saw a greater increase in revenue after the policy change. Second, we show that smaller volume, high quality sites benefited the most from the policy change. 

\subsection{Sites with thin markets experienced lower increases in prices}

Theoretically, the effect of context disclosure on site placement should be moderated by the competitiveness of the market (see Figures \ref{fig:sim_het_thick} and \ref{fig:sim_het_thin}). This motivates an investigation of heterogeneous treatment effects across sites with stronger or weaker \emph{competition} prior to the policy change. Figure \ref{fig:competitiveness} shows the average daily buyers for each site prior to the policy change. From this we create a dummy variable for sites that had fewer buyers (thin markets) prior to the policy change. We set the cutoff point at the first quartile, which is 28 average daily buyers. This variable captures market thinness before the policy change, and we assume that sites with thinner markets before the policy change were likely to have thinner markets after the change. Figure \ref{fig:competitiveness} indicates that this is a reasonable assumption since market competitiveness was similar before and after treatment. This pre-treatment covariate is also not contaminated by the treatment. Based on our simulations in Section \ref{sec:theory} and the literature on auctions, we expect sites with thick markets to experience a greater increase in average revenue per impression, because it is more likely that there are several buyers who will value those impressions more when provided with context information.
\begin{table}[!htbp] 
\centering 
\caption{Heterogeneous treatment effects for sites with thin markets - Diff-in-diff analysis of the change in average revenue per impression (\euro{} per thousand) due to context disclosure.} 
\label{treatment_effects1} 
\small 
\begin{threeparttable}
\begin{tabular}{@{\extracolsep{8pt}}l D{.}{.}{7} D{.}{.}{7} } 
\hline 
\hline \\[-1.5ex]
& \multicolumn{2}{c}{Average revenue} \\
& \multicolumn{2}{c}{ per impression} \\ 
\\[-1ex]
\hline 
\\[-1ex]
\textbf{Treatment effects}   \\ 
Partial disclosure x Year16 &  0.108^{***} & (0.032)\\ 
Partial disclosure x Year16 x Thin & 0.028  & (0.052) \\ 
[1ex]
Full disclosure x Year16 & 0.154^{***} & (0.039)  \\ 
Full disclosure x Year16 x Thin &  -0.150 & (0.092) \\  
[1ex]
\textbf{Baselines}  &  \\ 
[1ex]
Constant & 1.728^{***} & (0.195)\\ 
Partial disclosure & -0.006  & (0.020) \\ 
Full disclosure & 0.182^{***} & (0.035) \\ 
Year16 & 0.193^{**} & (0.027) \\ 
Thin & -0.110 & (0.132) \\ 
Partial disclosure x Thin & 0.056^{**} & (0.024) \\ 
Full disclosure x Thin & 0.069 & (0.046)  \\ 
Year16 x Thin & -0.298^{***}  & (0.038) \\ 
[1ex]
\textbf{Controls} &  \\ 
[1ex]
Supply in millions & -0.001^{**} & (0.0004) \\ 
Average daily buyers & -0.005^{***} & (0.001) \\ 
Monthly ad spending & 0.0002 & (0.0005)\\ 
[1ex]
\hline \\[-1.5ex]
Site FE  & \multicolumn{2}{c}{Yes} \\  
N & \multicolumn{2}{c}{3058}\\ 
Adjusted R$^{2}$ & \multicolumn{2}{c}{0.7718}  \\ 
[1ex]
\hline
\hline 
\end{tabular} 
\begin{tablenotes}
\item Notes: Standard errors, in parentheses, are clustered at the week level. $^{***}$Significant at the 1 percent level. $^{**}$Significant at the 5 percent level. $^{*} $Significant at the 10 percent level. As a robustness check these regressions were also run excluding the week in 2015 when prices spiked (calendar week 6, 2015) and substantive results remain the same.
\end{tablenotes}
\end{threeparttable}
\end{table} 

\begin{figure}[!ht]
\caption{Differential effect of context disclosure for sites with thin markets.}%
\label{fig:effect_plot_thick_thin}%
\begin{center}
\includegraphics[width=0.7\textwidth]{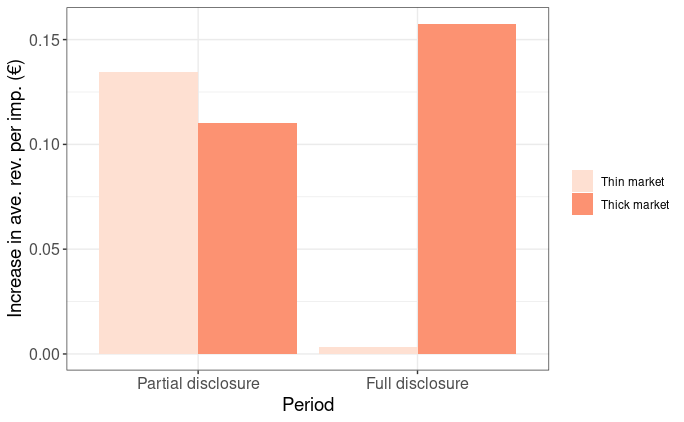}
\end{center}
\end{figure}

The model reported in Table \ref{treatment_effects1} shows the moderating effect of competition on average revenue per impression. The estimated increase in revenue per impression was 15.4 EUR cents for sites with thick markets (see \emph{Full disclosure x Year16}), while it was nearly zero for thin markets.\footnote{Note that sites with more buyers tend to sell more impressions, thus the effects sizes for thick markets here are very close the the overall averages from the volume-weighted regression reported in Table \ref{tab:main_result}.} The estimated effect for sites with thin markets is the sum of \emph{Full disclosure x Year16} and  \emph{Full disclosure x Year16 x Thin} which is 15.4 - 15.0 = 0.4 EUR cents. The finding that prices did not increase for sites with thinner markets is consistent with the literature on auctions and the simulations in Section \ref{sec:theory}, thus serving as an additional mechanism check. For ease of interpretation, we plot the estimated effects for sites with thin versus thick markets in Figure \ref{fig:effect_plot_thick_thin}. 

\subsection{Small, premium sites saw the greatest increase in prices}
For our final analysis, we look at heterogeneous treatment effects for sites of different quality and size. While slightly less theoretically motivated, it gives us an answer to the practical question of ``Which sites benefit the most from context transparency?" Even though we study a private exchange with generally brand-safe sites, not all of them are -- what advertisers may consider -- the highest-quality advertising outlets. To categorize sites according to their quality we asked three industry experts (a head of media planning, a media planner, and a trader in RTB auctions) to classify the sites into those that provide premium and non-premium advertising environments, which is a common industry categorization. All experts were familiar with the sites and had purchased media from the ad exchange in the past. In addition, we categorized sites into small versus large based on the number of impressions they provided to the RTB market.\footnote{Sites are categorized as ``large" if their daily supply of impressions on the RTB platform was larger than 80,000 in March 2016, otherwise they are coded as ``small." Considering the distribution of daily average supply of impressions, 80,000 daily impressions per site is a clear cut-off point which separates the sites with high and low supply of impressions.} That leads to four categories of sites: Premium large, Premium small, Non-premium large, and Non-premium small.

\begin{table}[!ht] 
\centering 
\caption{Heterogeneous treatment effects for high quality and large sites - Diff-in-diff analysis of the change in average revenue per impression (\euro{} per thousand) due to context disclosure.} 
\label{treatment_effects2} 
\small 
\begin{threeparttable}
\begin{tabular}{@{\extracolsep{5pt}}lD{.}{.}{-3} c}
\hline
\hline \\[-1.5ex] 
& \multicolumn{1}{c}{Average revenue} & \multicolumn{1}{c}{Std.}\\ 
& \multicolumn{1}{c}{per impression} & \multicolumn{1}{c}{Error}\\ 
\hline \\[-1.5ex] 
\textbf{Treatment effects}  &   \\ 
Partial disclosure x Year16 &  0.116^{***} & (0.032) \\
Partial disclosure x Year16 x Non-Premium+Large &  -0.023 & (0.037) \\ 
Partial disclosure x Year16 x Non-Premium+Small  & -0.014 & (0.069)\\ 
Partial disclosure x Year16 x Premium+Small  & -0.079 & (0.072) \\ 
[1ex]
Full disclosure x Year16  & 0.153^{***} & (0.044) \\ 
Full disclosure x Year16 x Non-Premium+Large & -0.021 & (0.063)\\ 
Full disclosure x Year16 x Non-Premium+Small   & 0.172^{*} & (0.091)\\ 
Full disclosure x Year16 x Premium+Small   &  0.316^{**} & (0.149)\\ 
[1ex]
\textbf{Baselines}  &  \\ 
[1ex]              
Constant  & -2.776^{***} & (0.321)\\ 
Partial disclosure  & 0.003 & (0.019)  \\ 
Full disclosure  & 0.164^{***} & (0.038) \\ 
Year16 & 0.208^{***} & (0.028) \\ 
Non-Premium+Large & 0.177^{***} & (0.061) \\ 
Non-Premium+Small & 2.428 & (0.212) \\ 
Premium+Small & 2.439^{***} & (0.213) \\ 
Partial disclosure x Non-premium large   & -0.052^{*} & (0.029) \\ 
Partial disclosure x Non-premium small   &  0.002 & (0.055)\\ 
Partial disclosure x Premium small   & -0.032 & (0.048)\\ 
Full disclosure x Non-premium large   & 0.082^{**} & (0.041)\\ 
Full disclosure x Non-premium small   &  -0.067 & (0.077)\\ 
Full disclosure x Premium small   & 0.224^{***} & (0.070) \\ 
Year16 x Non-premium large   & -0.047 & (0.030)  \\ 
Year16 x Non-premium small   & -0.094 & (0.058)  \\ 
Year16 x Premium small & -0.298^{***} & (0.056) \\
[1ex]
\textbf{Controls} &    \\ 
Supply in millions  & -0.001^{**} & (0.0004)  \\ 
Daily average buyers & 0.022^{***} & (0.002)  \\ 
Monthly ad spending & 0.0003 & (0.0005) \\ 
[1ex]
\hline \\[-1.5ex] 
Site FE  & \multicolumn{2}{c}{Yes} \\  
N & \multicolumn{2}{c}{3058} \\ 
Adjusted R$^{2}$ & \multicolumn{2}{c}{0.7787} \\ 
\hline 
\hline \\[-1.8ex] 
\end{tabular} 
\begin{tablenotes}
\small
\item Notes: Standard errors, in parentheses, are clustered at the week level. Baseline is January- March in 2015 and sites are premium and large in March 2015 for column (2). $^{***}$Significant at the 1 percent level. $^{**}$Significant at the 5 percent level. $^{*} $Significant at the 10 percent level. A robustness check filtering out calendar week 6, 2015 when prices spiked showed substantively similar results.
\end{tablenotes}
\end{threeparttable}
\end{table} 

\begin{figure}[!ht]
\caption{Differential effect of context disclosure for sites of varying quality and size.}
\label{fig:effect_plot_qual_size}%
\begin{center}
\includegraphics[width=0.7\textwidth]{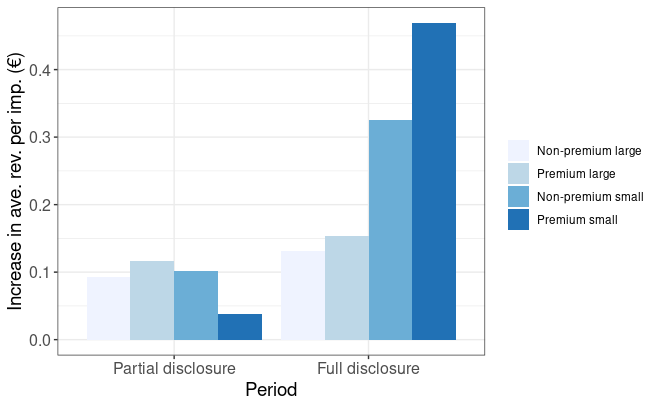}
\end{center}
\end{figure}

The model reported in Table \ref{treatment_effects2} shows the moderating effect of our size/quality-measure on average revenue per impression (again with controls and fixed effects). The effect of the policy change for these Premium large sites is shown in rows \emph{Partial disclosure x Year16} and \emph{Full disclosure x Year16}; rows \emph{Partial disclosure x Year16 x Non-premium large} and \emph{Full disclosure x Year16 x Non-premium large} show the incremental effect for non-premium/large sites. Similarly, the effects for Non-premium small and Premium small sites are shown in the table. Because the total effect for all four types of sites is difficult to determine from the regression table, we illustrate the estimated effects in Figure \ref{fig:effect_plot_qual_size}. Premium small sites benefit the most from  the policy change with an estimated effect more than three times that of the average (15.3 + 31.6 = 46.9 EUR cents). These small, premium sites typically serve a ``niche audience" with a very specific topical interest (e.g., a website that provides content targeted at physicians) and once ad buyers know what they are bidding for, some of them value these sites far more than had they been in a bundle of unknown sites. Thus, the data suggests that such sites have the most to gain by increasing context transparency.

\section{Discussion and implications}
\label{discussion}
This study investigated a specific change in context information where an exchange moved from providing only the level of the ``channel" to providing the subdomain associated with each ad impression in the bid request. Our analysis of the policy change shows that buyers value context information above-and-beyond user information and act on it as soon as context information is available. Consequently, average revenue per impression rose after the policy change relative to the previous year. As we illustrate with a simulation reported in Figure \ref{fig:sim_het_thick}, these effects are consistent with a scenario where ad buyers prefer different sites. Such heterogeneous preferences lead to an increase in prices with context disclosure \citep{tadelis2015information}, so long as the market does not become thin \citep{levin2010online, hummel2016does}. Under this scenario, ad buyers bid higher for the sites they each prefer after the policy change.

Several mechanism checks provide convergent evidence that this is indeed what happened: 1) markets thinned out slightly suggesting fewer bidders were bidding on each site, but remained sufficiently thick for prices to rise; 2) the distribution of winning bids shifted to the right meaning that winning bids were more dispersed when ad buyers were provided with context information; 3) most individual sites saw an increase in average revenue per impression; 4) consistent with our simulations in \ref{fig:sim_partial}, partial disclosure of information to a single buyer resulted in more auctions being won by this buyer. We are also able to rule out some alternative explanations. For one, the change in average revenue per impression was not due to ad buyers shifting budgets from programmatic direct to RTB. In addition, ad buyer and site turnover was not responsible for the observed increase in revenue per impression. Our evidence therefore points towards buyers increasing their bids in response to the policy change.  

Finally, to answer the practical question of whether certain types of sites benefited more from the policy change, we also investigate heterogeneous treatments effects: The increase in average revenue per impression was most pronounced for sites with a large number of average daily buyers prior to the policy change (again, consistent with our simulations in \ref{fig:sim_het_thick}). From a more managerial standpoint, we show that small, premium sites benefited the most from context disclosure. They typically serve a ``niche audience" with a very specific topical interest and once ad buyers know what they are bidding for, some of them value these sites far more than had they been in a bundle of unknown sites. Yet, there are almost no losers of the policy change in this market -- we mainly see sites that benefit more and sites that benefit less. 

These changes in prices were economically meaningful for this private exchange. The average weekly supply for a site in our sample is roughly 3.5M impressions (see Table \ref{table:ss_supply}), sold for an average CPM of 88 EUR cents. According to our analysis, average CPM rises by about 15.4 EUR cents when all ad buyers are provided with placement information (see Table \ref{tab:main_result}, Column 1, Full disclosure x Year16). Therefore, on average each site generates an additional yearly revenue of (3.5M/1000) x 15.8 EUR cents x 52 weeks = \euro{28,756}. Hence, the overall revenue across all sites increases by about \euro{28,756} x 57 = \euro{1,639,092} per year. Since ad exchanges typically receive a revenue share from the sites in their portfolio (in our case about 2.5\%), the additional yearly revenue for our ad exchange is roughly \euro{1,639,092} x 2.5\% = \euro{40,977}. Note that this calculation depends on the figures obtained from our sample and is highly dependent on the scale of the ad exchange and the sites in its portfolio. In addition, our data does not allow us to investigate whether the effect might vanish in the future. Yet, we see no reason why the effect would not continue. Therefore, these rough calculations show that the policy change is associated with a substantial increase in revenue for the ad exchange and its publishers. 

Although we demonstrate the impact of information disclosure on average revenue per impression utilizing a data set that consists of winning auction outcomes, we would be able to gain more insight into buyer's valuations by investigating data on individual ad buyers' bids for specific impressions. Instead, we only have data on the selling prices for the winning bids, which makes it difficult to determine precisely how the ad placement information affected each bidder's valuation. In addition, if we had data on individual bids, we could better assess market competitiveness by counting the number of ad buyers bidding for each impression. Despite these limitations, our study represents an important step in empirically understanding the effect of placement information in advertising auctions. Moreover, our results have important implications for different market players in digital marketing. 

\subsection{Implications}
The extensibility of our findings to other markets depends critically on the sites and ad buyers that participate in the market. Some of the sites participating in the exchange we study provide niche content, but even the smaller websites were brand-safe, reputable media outlets that had been vetted by the exchange. This mix of sites is typical of private exchanges, and so our results strongly suggest that private exchanges should provide information at the subdomain in the bid request. While we do not see buyer entry due to the policy change within our observation window (three months post-disclosure), it might be possible that buyers will want to join this specific private market if others do not provide the same level of context disclosure. Such buyer entry might increase auction revenue even further. 

While our results translate fairly directly to other private exchanges, which are a growing share of the display advertising market \citep{emarketer2020}, is is more difficult to say how context disclosure might affect open exchanges. Open markets attract a much wider range of sites (including ``clickbait" websites, fake news, and other sites with low quality advertising environments) and it is currently left to individual publishers to decide whether to \textit{truthfully} disclose the context for ad impressions they sell. There is certainly a higher risk of deconflation for sites that produce extreme content and some may become ``orphaned" by ad buyers, as evidenced by the recent drop in demand for advertising at the alt-right site Breitbart.com in the US \citep{wapo18} when buyers became aware that their programmatic ads were appearing on the site. Increased context transparency may force them to leave the exchange if their revenues decrease substantially over time. Yet efforts by the IAB to increase transparency in the open markets \citep{domainspoof} will, if successful, improve outcomes for ad buyers, reputable publishers, exchange operators, and the industry as a whole. 

It also means that reliable publishers may want to avoid selling impressions on the open exchanges when they are not transparent enough. Instead they may want to form their own private markets as they can expect auction revenues to rise when providing more context transparency. Some publishers in the Netherlands have already done so, eliminating user tracking all together \citep{edelman2020}. Interestingly, digital revenue rose for those publishers. Private exchanges are increasingly popular and are expected to process the majority of display advertising impressions in 2020 \citep{emarketer2020}. 

Our analysis shows that site placement information provides ad buyers with additional information about the value of an impression, above-and-beyond the rich cookie information available to European ad buyers in 2016. That is, context information is complementary to user information. If context information is also a partial substitute for user information, then context information will become even more important as ad buyers' access to user information becomes more limited. Regulations like GDPR already limit the amount of user-level targeting that is possible. In addition, Google recently announced efforts to limit the use of third-party cookies in their Chrome browser by making “disable third-party cookies” the standard setting \citep{chromecookies1, chromecoookies2}. Thereby, the market share of browsers (including Mozilla, Safari, and  Chrome) that inhibit tracking will grow to more than 80\% in many countries in the next two years. Thus, user targeting will be less straightforward in the future and industry experts predict that contextual targeting will become more relevant \citep{chromecoookies3, chromecoookies4}.   

There are few empirical studies that investigate the effect of reduced access to user information on outcomes of ad auctions. A recent working paper by \citet{abhishek2019Tracking} shows that when the user's cookie is available, publisher's revenue increases, but the increase is just about 4\% corresponding to an average increase of \$0.00008 per ad. The authors argue that the information technology required to track user behavior along with organizational measures that ensure compliance with privacy regulations come at a cost and sometimes a prohibitive one, making it unattractive for publishers to enable cookie tracking. This seems to be born out in the industry as publishers like \emph{The New York Times} have moved to protect users by decreasing the amount of user information collected on their sites \citep{berjon2020}. In contrast, providing ad buyers with more context information is nearly cost less to publishers; we provide convergent evidence that doing so results in a substantial revenue increases.  

While it is speculating beyond our data, we expect full URL disclosure to be very attractive to ad buyers who can then target ads against the specific content in an article. In fact, startups such as Grapeshot (acquired by Oracle in 2018), Peer39, and Leiki (acquired by DoubleVerify) have been building machine learning tools to help ad buyers determine which URLs are most attractive, based on the text on the page. However, finer levels of context disclosure may lead to thin markets and deconflation for specific URLs or particular content topics. At the same time, it may encourage content creators and publishers to focus on content that is appealing to consumers and ad buyers rather than simply attracting an audience to generate impressions \citep{mediabias}. We encourage future theoretical research that investigates how such fine-grained context disclosure could impact publishers' incentives to produce content and the welfare of advertisers, publishers, and content consumers.

\clearpage
\linespread{1.1}
\bibliographystyle{chicago} 
\bibliography{main} 
\linespread{1.5}

\newpage
\appendix
\noindent \Large{\textbf{Web Appendix}}

\renewcommand\thefigure{A.\arabic{figure}}
\renewcommand\thetable{A.\arabic{table}}

\section{Interface of demand-side platform}
\setcounter{figure}{0} 
\setcounter{table}{0}
\label{interface}

\begin{figure}[ht]
\caption{Example of a demand-side platform (DSP) interface where users can enter the URLs of the sites where they want their ads to appear.}
\label{fig:url}
\centering
\includegraphics[width=6in]{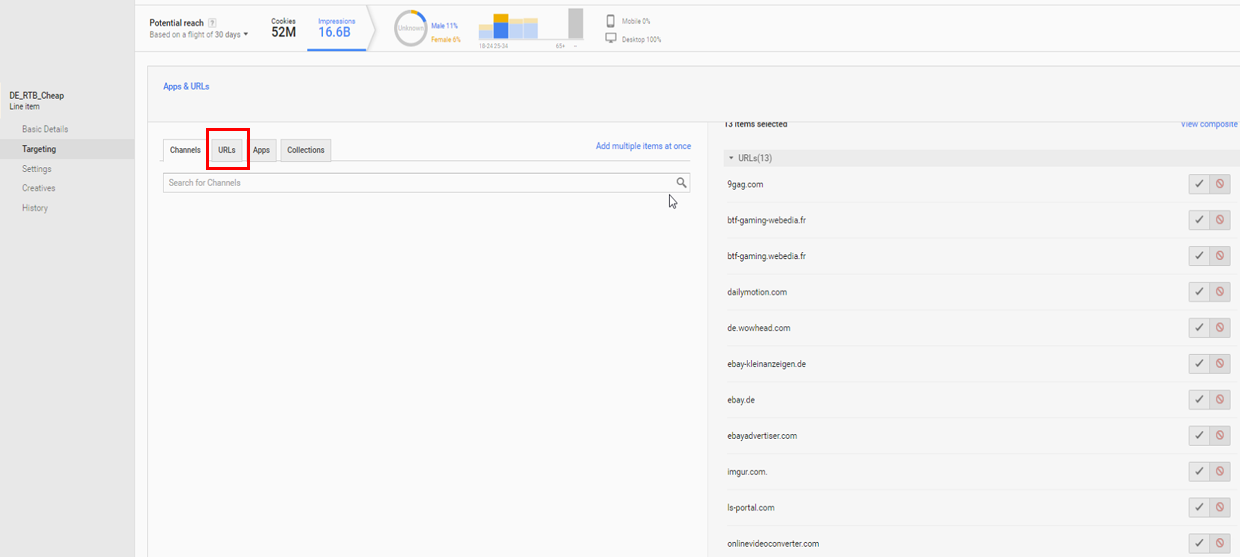}
\end{figure}

\newpage
\section{Effects of context disclosure by month}
\label{sec:monthlyeffects}

\begin{table}[!htbp] 
\caption{Treatment effects month by month.} 
\label{tab:monthly_diff_in_diff} 
\centering
\begin{threeparttable}
\begin{tabular}{@{\extracolsep{1pt}}l D{.}{.}{-3} D{.}{.}{-3} } 
\hline 
\hline  
& \multicolumn{1}{c}{Average revenue} & \multicolumn{1}{c}{Average revenue}  \\ 
& \multicolumn{1}{c}{per impression} & \multicolumn{1}{c}{per impression}\\
\hline
\textbf{Treatment effects} \\
April (Partial disclosure) x Year16 & 0.108^{***} & 0.099^{***} \\
& (0.032) & (0.028)\\ 
May (Full disclosure) x Year16 & 0.206^{***} & 0.187^{***} \\ 
& (0.037) & (0.038) \\ 
June (Full disclosure) x Year16 & 0.073^{*} &  0.025 \\
& (0.039) & (0.041) \\ 
July (Full disclosure) x Year16 & 0.125^{***} & 0.079^{*} \\
& (0.049) & (0.051)  \\
\textbf{Baselines}\\
Constant (No disclosure) & 1.671^{***} & 0.933^{***} \\ 
& (0.189) & (0.057) \\ 
April (Partial disclosure) & -0.007 &  0.016 \\
& (0.019) & (0.022)\\ 
May (Full disclosure) & 0.105^{***} &  0.151^{***} \\
& (0.035) & (0.042) \\ 
June (Full disclosure) & 0.247^{***} & 0.320^{***} \\
& (0.019) & (0.023) \\
July (Full disclosure) & 0.243^{***} & 0.287^{***}\\
& (0.040) & (0.035) \\ 
Year16 & 0.193^{***} & 0.234*** \\
& (0.027) & (0.027) \\ 
\textbf{Controls} \\
Supply in millions & -0.001^{***} & \\
& (0.0003) \\ 
Daily average buyers & 0.005^{***} & \\
& (0.001) \\ 
Monthly ad spending & 0.0004 & \\
& (0.001) \\ 
\hline
Site FE & \multicolumn{1}{c}{Yes} & \multicolumn{1}{c}{Yes} \\ 
N & \multicolumn{1}{c}{3058} & \multicolumn{1}{c}{3058}\\
Adjusted $R^2$ & \multicolumn{1}{c}{0.7785} & \multicolumn{1}{c}{0.7716} \\
\hline 
\hline \\[-1.8ex] 
\end{tabular} 
\begin{tablenotes}
\small
\item Notes: Standard errors, in parentheses, are clustered at the week level. $^{***}$Significant at the 1 percent level. $^{**}$Significant at the 5 percent level. $^{*} $Significant at the 10 percent level. A robustness check filtering out the spikes in price in calendar week 6, 2015 showed substantively similar results.
\end{tablenotes}
\end{threeparttable}
\end{table} 

\newpage
\section{Distribution of winning bids in 2015}
\begin{figure}[htbp!]
\centering
\includegraphics[width=\textwidth]{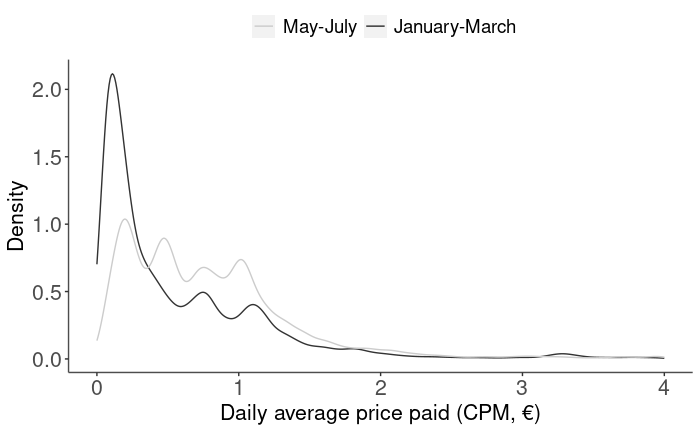}
\caption{Density plot of average price paid per impression for 2015.}
\label{fig:dist_price_2015}
\end{figure}

\newpage
\section{Placebo tests for synthetic control}
\label{sec:placebo} 
\begin{figure}[ht]
\caption{Distribution of weekly estimated gaps for treated and control buyers.} \label{fig:gaps}%

\subfloat[Average winning price gaps for treated buyer (thick black line) and placebo gaps for buyers in control group (grey lines).]{%
\includegraphics[width=0.9\textwidth]{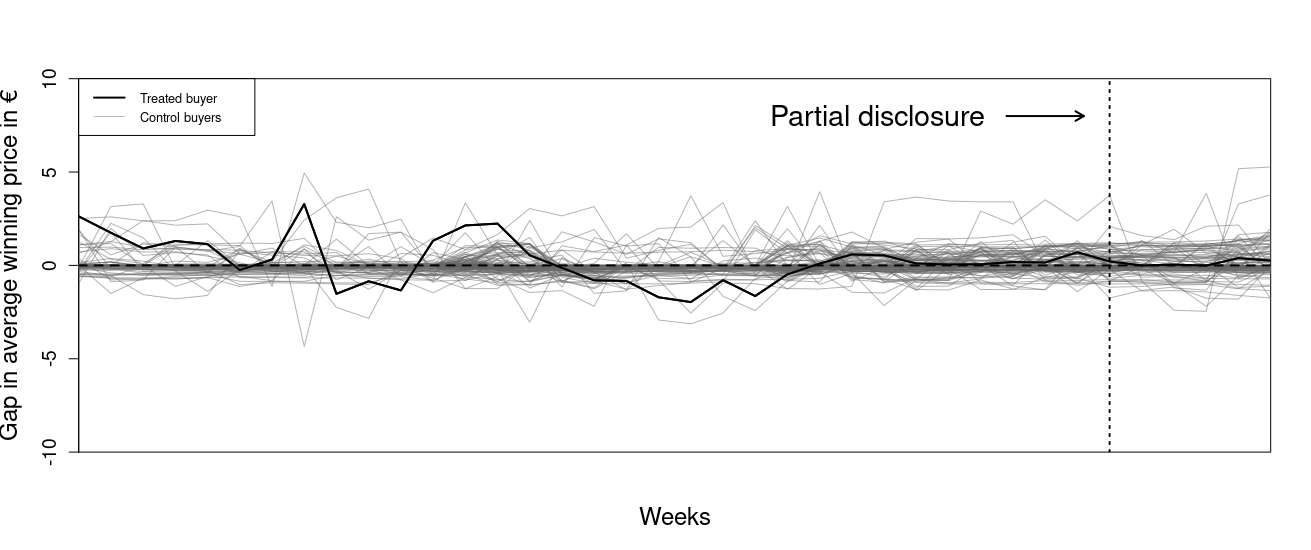}
\label{synth_price_gap}
}
\qquad
\subfloat[Number of impressions won gaps for treated buyer (thick black line) and placebo gaps for buyers in control group (grey lines).]{%
\includegraphics[width=0.9\textwidth]{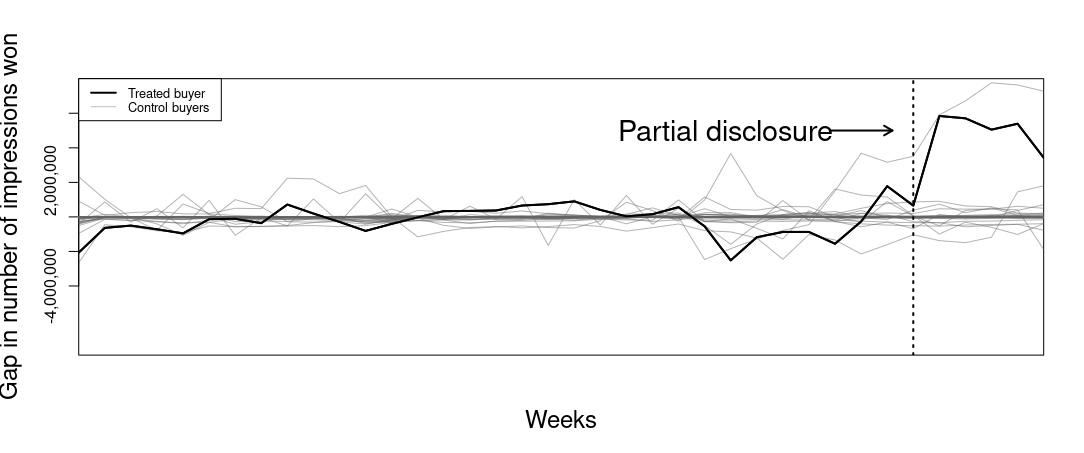}
\label{synth_imps_gap}
}
\end{figure}

\normalsize
To assess statistical significance of the synthetic control results, we conduct a series of placebo tests by applying the synthetic control method to the ad buyers who were not provided with placement information.\footnote{This is the standard method of assessing significance for synthetic controls \cite[cf.][]{tirunillai2017does}.} By doing so, we produce a distribution of weekly estimated gaps between each ad buyer and its optimal synthetic control (see Figure \ref{fig:gaps}). The quality of fit of the synthetic control can be assessed by using the mean squared prediction error (MSPE) prior to the policy change. Following \citet{abadie2003economic, abadie2010synthetic} and \citet{tirunillai2017does}, Figure \ref{fig:gaps} visualizes the placebo buyers having a pre-intervention MSPE of less than 5 times the MSPE of the treated buyer which results in 131 control buyers in Figure \ref{synth_price_gap} and 121 control buyers in Figure \ref{synth_imps_gap}. As denoted by the thick black lines in Figure \ref{synth_price_gap} and \ref{synth_imps_gap}, the synthetic control method provides a very good fit for the treated buyers. The estimated number of impressions won has a p-value of 0.016.\footnote{p-values are calculated by means of the ratio of post - pre intervention MSPE. If an ad buyer were randomly treated, the probability of obtaining a post - pre intervention MSPE ratio as large as the one for the treated buyer would be 2 (number of ad buyers exceeding the treated buyer's MSPE ratio) over 127 (number of ad buyers).}, suggesting the theoretically-predicted, significant increase in impressions won for the treated bidder. Also consistent with theory, we do not observe a significant effect on average winning price in the same period (p-value = 0.932). 

\end{document}